\documentclass[aps,prd,twocolumn,superscriptaddress,nofootinbib,preprintnumbers]{revtex4-1}
\def\aj{AJ}%
\def\apj{ApJ}%
\def\apjs{ApJS}%
\def\aap{A\&A}%
\def\mnras{MNRAS}%
\def\prd{Phys.~Rev.~D}%
\def\pasp{PASP}%
\def\pasj{PASJ}%
\def\jcap{JCAP}%
\def\physrep{Phys.~Rep.}%
\def\3x2pt{3$\times$2pt}
\def\5x2pt{5$\times$2pt}
\def\6x2pt{6$\times$2pt}

\newcommand{\planck}{\textit{Planck}}

\newcommand{\metacal}{\textsc{Metacalibration}}

\newcommand{\redmagic}{\textsc{redMaGiC}}
\newcommand{\maglim}{\textsc{MagLim}}

\usepackage[utf8]{inputenc}
\usepackage[T1]{fontenc}
\usepackage{graphicx}
\usepackage{dcolumn}
\usepackage{bm}
\usepackage{hyperref}
\usepackage{amsmath}	
\usepackage{amssymb}	
\usepackage{slashed}
\usepackage[normalem]{ulem}
\usepackage[dvipsnames]{xcolor}
\usepackage{xspace}
\usepackage{multirow}
\usepackage{booktabs}
\usepackage{comment}
\usepackage{lineno}
\RequirePackage{ifxetex}

\defcitealias{y3-nkgkmethods}{\textsc{Paper I}}
\defcitealias{y3-nkgkmeasurement}{\textsc{Paper II}}
\defcitealias{y3-5x2kp}{\textsc{Paper III}}

\newcommand{\be}{\begin{equation}}
\newcommand{\ee}{\end{equation}}
\newcommand{\ba}{\begin{eqnarray}}
\newcommand{\ea}{\end{eqnarray}}

\newcommand{\nside}{\ifmmode N_{\mathrm{side}}\else $N_{\mathrm{side}}$\fi}
\newcommand{\npix}{\ifmmode n_{\mathrm{pix}}\else $n_{\mathrm{pix}}$\fi}
\newcommand{\Npix}{\ifmmode N_{\mathrm{pix}}\else $n_{\mathrm{pix}}$\fi}
\newcommand{\lmin}{\ifmmode \ell_{\mathrm{min}}\else $\ell_{\mathrm{min}}$\fi}
\newcommand{\lmax}{\ifmmode \ell_{\mathrm{max}}\else $\ell_{\mathrm{max}}$\fi}

\usepackage[utf8]{inputenc}
\usepackage{newunicodechar,graphicx}
\DeclareRobustCommand{\okina}{%
  \raisebox{\dimexpr\fontcharht\font`A-\height}{%
    \scalebox{0.8}{`}%
  }%
}
\newunicodechar{ʻ}{\okina}

\defcitealias{5x2methods}{B18}
\defcitealias{NKpaper}{O18a}
\defcitealias{GKpaper}{O18b}
\defcitealias{Krause:2017}{K17}
\defcitealias{Melchior:2017}{M17}
\defcitealias{Omori:2017}{O17}
\defcitealias{DESy1:2017}{DES-Y1-3x2}

\newcommand{\galshear}{$\langle \delta_{\rm g} \gamma_{\rm t}\rangle$}
\newcommand{\shearshear}{$\langle \gamma \gamma \rangle$}

\newcommand{\nn}{$\langle \delta_{\rm g} \delta_{\rm g} \rangle$}
\newcommand{\nk}{$\langle \delta_{\rm g} \kappa_{\rm CMB}\rangle $}
\newcommand{\kapkap}{$\langle \kappa_{\rm CMB} \kappa_{\rm CMB}\rangle $}
\newcommand{\gk}{$\langle \gamma_{\rm t}\kappa_{\rm CMB}\rangle$}
\newcommand{\kk}{$\langle\kappa_{\rm CMB} \kappa_{\rm CMB}\rangle$}

\newcommand{\fivetwo}{$5$$\times$$2{\rm pt}$}
\newcommand{\threetwo}{$3$$\times$$2$${\rm pt}$}
\newcommand{\otherthreetwo}{$\overline{3\!\times\! 2 {\rm pt}}$}

\newcommand{\sixtwo}{$6$$\times$$2{\rm pt}$}
\newcommand{\nkgk}{$\langle \delta_g \kappa_{\rm CMB}\rangle + \langle \gamma_{\rm t}\kappa_{\rm CMB}\rangle$}

\begin{document}
\title[Joint analysis of DES Year 3 data and CMB lensing from SPT and Planck III: Cosmological constraints]
{Joint analysis of DES Year 3 data and CMB lensing from SPT and Planck III:\\ Combined cosmological constraints}

\author{T.~M.~C.~Abbott}
\affiliation{Cerro Tololo Inter-American Observatory, NSF's National Optical-Infrared Astronomy Research Laboratory, Casilla 603, La Serena, Chile}
\author{M.~Aguena}
\affiliation{Laborat\'orio Interinstitucional de e-Astronomia - LIneA, Rua Gal. Jos\'e Cristino 77, Rio de Janeiro, RJ - 20921-400, Brazil}
\author{A.~Alarcon}
\affiliation{Argonne National Laboratory, 9700 South Cass Avenue, Lemont, IL 60439, USA}
\author{O.~Alves}
\affiliation{Department of Physics, University of Michigan, Ann Arbor, MI 48109, USA}
\affiliation{Laborat\'orio Interinstitucional de e-Astronomia - LIneA, Rua Gal. Jos\'e Cristino 77, Rio de Janeiro, RJ - 20921-400, Brazil}
\author{A.~Amon}
\affiliation{Institute of Astronomy, University of Cambridge, Madingley Road, Cambridge CB3 0HA, UK}
\affiliation{Kavli Institute for Cosmology, University of Cambridge, Madingley Road, Cambridge CB3 0HA, UK}
\author{F.~Andrade-Oliveira}
\affiliation{Department of Physics, University of Michigan, Ann Arbor, MI 48109, USA}
\author{J.~Annis}
\affiliation{Fermi National Accelerator Laboratory, P. O. Box 500, Batavia, IL 60510, USA}
\author{B.~Ansarinejad}
\affiliation{School of Physics, University of Melbourne, Parkville, VIC 3010, Australia}
\author{S.~Avila}
\affiliation{Instituto de Fisica Teorica UAM/CSIC, Universidad Autonoma de Madrid, 28049 Madrid, Spain}
\author{D.~Bacon}
\affiliation{Institute of Cosmology and Gravitation, University of Portsmouth, Portsmouth, PO1 3FX, UK}
% ...

\collaboration{The DES and SPT Collaborations}

\date{Last updated \today}

\label{firstpage}


\begin{abstract}
We present cosmological constraints from the analysis of two-point correlation functions between galaxy positions and galaxy lensing measured in Dark Energy Survey (DES) Year 3 data and measurements of cosmic microwave background (CMB) lensing from the South Pole Telescope (SPT) and {\it Planck}.  When jointly analyzing the DES-only two-point functions and the DES cross-correlations with SPT+{\it Planck} CMB lensing, we find $\Omega_{\rm m} = 0.344\pm 0.030$ and $S_8 \equiv \sigma_8 (\Omega_{\rm m}/0.3)^{0.5} = 0.773\pm 0.016$, assuming $\Lambda$CDM.  When additionally combining with measurements of the CMB lensing autospectrum, we find $\Omega_{\rm m} = 0.306^{+0.018}_{-0.021}$ and $S_8  =  0.792\pm 0.012$.  The high signal-to-noise of the CMB lensing  cross-correlations enables several powerful consistency tests of these results, including  comparisons with constraints derived from cross-correlations only, and comparisons designed to test the robustness of the galaxy lensing and clustering measurements from DES.  Applying these tests to our measurements, we find no evidence of significant biases in the baseline cosmological constraints from the DES-only analyses or from the joint analyses with CMB lensing cross-correlations. However, the CMB lensing cross-correlations suggest possible problems with the correlation function measurements using alternative lens galaxy samples, in particular the \redmagic{} galaxies and high-redshift \maglim{} galaxies, consistent with the findings of previous studies.  We use the CMB lensing cross-correlations to identify directions for further investigating these problems.
\end{abstract}

\preprint{DES-2021-0649}
\preprint{FERMILAB-PUB-22-475-PPD}

\maketitle

\section{Introduction}
\label{sec:intro}

The late-time large scale structure (LSS) of the Universe is sensitive to a variety of cosmological signals, ranging from the properties of dark energy and dark matter, to the masses of the neutrinos.  Galaxy imaging surveys probe this structure using observations of both the positions of galaxies (which trace the LSS) and the shapes of galaxies (which are distorted by the gravitational lensing effects of the LSS).  Several galaxy imaging surveys have used two-point correlations between these measurements to place constraints on cosmological models, including the Kilo Degree Survey (KiDS), the Hyper Suprime Cam Subaru Strategic Program (HSC-SSP), and the Dark Energy Survey (DES)  \citep[e.g.][]{Heymans:2021,Hamana:2020,y13x2}.  DES has recently presented cosmological constraints from the joint analysis of the three two-point correlation functions (\threetwo{}) between measurements of these probes from the first three years (Y3) of DES data \cite{y3-3x2ptkp}.

Surveys of the cosmic microwave background (CMB) are also able to probe the late-time LSS through the effects of gravitational lensing.  Although CMB photons originate from the last scattering surface at redshift $z \sim 1100$, their paths are perturbed by structure at late times, including the same LSS measured by galaxy surveys.   CMB lensing provides a highly complementary probe of structure to galaxy surveys, and cross-correlations between the two have several appealing features.  For one, current galaxy imaging surveys (like DES) identify galaxies out to $z \sim 1$, but the galaxy lensing measurements with these surveys do not have significant sensitivity beyond $z \simeq 0.75$.  Without the high-redshift lensing information, cosmological constraints from galaxy surveys at $z \gtrsim 0.75$ are therefore significantly degraded.  CMB lensing, however, reaches peak sensitivity at $z \sim 2$.  Therefore, by cross-correlating galaxy surveys with CMB lensing measurements, it is possible to obtain high-precision measurements of the evolution of the matter distribution over a broader range of redshifts than by using galaxy surveys alone.  Cross-correlations of galaxy surveys with CMB lensing measurements are also expected to be robust to certain types of systematic biases.  Because the galaxy survey measurements are so different from the CMB lensing measurements (e.g., they use data measured by different telescopes at different wavelengths, and use different estimators for the lensing signal), biases in the galaxy surveys are unlikely to correlate with biases in CMB lensing, making two-point functions between the two especially robust.  
Finally, cross-survey correlations often have different parameter dependencies than correlations within a survey, offering the possibility of improved parameter constraints via degeneracy breaking in joint analyses.

The prospect of obtaining tighter and more robust cosmological constraints from the late-time matter distribution via cross-correlations is particularly timely given recent hints of tensions between some cosmological probes.  In particular, recent observations of late-time structure from galaxy surveys tend to prefer lower values of $S_8 \equiv \sigma_8 \sqrt{\Omega_{\rm m}/0.3}$ than CMB surveys \cite{Battye:2015,MacCrann:2015,Raveri:2016,Raveri:2018,Planck:cosmo}.  This tension could result from physics beyond the standard cosmological constant and cold dark matter model ($\Lambda$CDM), or it could result from systematic biases in the analyses.  By cross-correlating galaxy surveys with CMB lensing, we obtain an independent handle on the late-time large scale structure measurements that can be used to investigate the origins of this possible tension \cite{Krolewski:2021,Robertson2021,y3-nkgkmeasurement,White:2022}.  Recent analyses have also suggested the possibility of systematic biases in galaxy survey measurements \cite{Chang:2019}.  Because cross-correlations between galaxy surveys and CMB lensing are robust to many important sources of systematic error, they provide a powerful way to ensure that late-time measurements of structure are unbiased.

This work presents the joint cosmological analysis of two-point correlations between galaxy positions and galaxy lensing measured in DES data, and CMB lensing measurements from the South Pole Telescope \citep[SPT,][]{SPT} and the {\it Planck} satellite \citep{planck2011a}.  As part of its 2008-2011 SPT-SZ survey, SPT obtained high-resolution and high-sensitivity maps of the CMB that partially overlap with the full DES footprint \citep{Omori:2017}.  At somewhat lower sensitivity and resolution, {\it Planck} has obtained full-sky maps of the CMB that overlap completely with the DES footprint.  Together, these CMB maps enable high signal-to-noise estimation of the CMB lensing signal across the entire DES footprint \cite{Omori:2017,y3-nkgkmethods}, presenting an opportunity for cross-correlation studies.

From the measurements of galaxy positions (used to compute the galaxy overdensity, $\delta_{\rm g}$), galaxy lensing ($\gamma$, or $\gamma_{\rm t}$ for the tangential shear), and CMB lensing ($\kappa_{\rm CMB}$), it is possible to form six two-point functions: galaxy clustering ($\langle \delta_{\rm g}\delta_{\rm g} \rangle$), galaxy-galaxy lensing ($\langle \delta_{\rm g}\gamma \rangle$), cosmic shear ($\langle \gamma\gamma \rangle$), galaxy density-CMB lensing cross-correlation (\nk{}), galaxy shear-CMB lensing cross-correlation ($\langle \gamma\kappa_{\rm CMB} \rangle$), and the CMB lensing auto-correlation ($\langle \kappa_{\rm CMB}\kappa_{\rm CMB} \rangle$). All six of the above will be considered here (hereafter, we refer to this combination as \sixtwo{}).  The five two-point functions excluding the CMB lensing auto-correlation (referred to as \fivetwo{}) all probe structure below about at $z \lesssim 1.25$, and are highly correlated.  This combination, which we measure using DES, SPT and {\it Planck} data is the primary focus of this work.  The CMB lensing autocorrelation measurements used in this analysis are derived from all-sky {\it Planck} data, and are minimally correlated with the \fivetwo{} measurements owing to their small (fractional) sky overlap and sensitivity to higher redshifts \cite{y3-nkgkmethods}.  We therefore treat the CMB lensing autocorrelation as an external probe, and combine it with \fivetwo{} at the likelihood level.

As highlighted above, one of the key reasons to consider cross-correlations of galaxy surveys with CMB lensing is to improve robustness to systematic uncertainties.  We will therefore also analyze various subsets of the \sixtwo{} probes for the purposes of testing robustness and exploring sensitivity to possible systematic errors.  Of particular interest for these tests is the unexpected discovery in DES Y3 data of discrepancies in the galaxy bias values preferred by the clustering and lensing measurements.  The DES Y3 analysis considered two galaxy samples for the purposes of measuring $\delta_g$: \maglim{} and \redmagic{}.  The \maglim{} galaxies at $z \lesssim 0.8$ were used for the baseline cosmological results presented in \cite{y3-3x2ptkp}.  Surprisingly, the galaxy bias values inferred for \redmagic{} galaxies from their clustering were found to be roughly 10\% lower than the bias values inferred from lensing \cite{y3-2x2ptbiasmodelling}, with this discrepancy increasing for the highest-redshift galaxies.  There is no known physical explanation for this discrepancy, but tests in \cite{y3-2x2ptbiasmodelling} suggest that it may be connected to observational systematics imparting additional clustering power.  \maglim{} galaxies at high redshift ($z \gtrsim 0.8$) also showed a discrepancy between clustering and lensing \cite{y3-2x2ptaltlensresults}.  Further investigating these discrepancies is one of the main goals of the present analysis.

The analysis presented here makes several significant improvements relative to previous cross-correlation analyses between DES and SPT/{\it Planck} measurements of CMB lensing \cite{Baxter:2016,kirk16,giannantonio16,Baxter2018,Omori19a,Omori19b,5x2Y1}.  First, the DES data have significantly expanded in going from Y1 observations to Y3, covering roughly a factor of three larger area. Second, the CMB lensing maps from SPT/{\it Planck} have been remade with several improvements (described in more detail in \cite{y3-nkgkmethods}). Foremost among these is that we have used the CMB lensing estimator from \cite{madhavacheril18} to reduce contamination in the lensing maps from the thermal Sunyaev-Zel'dovich (tSZ) effect.  This contamination was the dominant source of systematic uncertainty for the analysis of \citep{5x2Y1}, and required us to remove a significant fraction of the small-scale measurements from our analysis to ensure that our results were unbiased.  As a result, the total signal-to-noise of the CMB lensing cross-correlations was significantly reduced. Using a CMB lensing map that is immune to contamination from the tSZ effect allows us to extract signal from a wider range of angular scales and hence improve our signal-to-noise ratio.  Finally, we have also implemented several improvements to the modeling of the correlation functions, which are described in more detail in \cite{y3-nkgkmethods}.

The analysis presented here is the last in a series of three papers: In \citep[][hereafter \citetalias{y3-nkgkmethods}]{y3-nkgkmethods} we described the construction of the combined, tSZ-cleaned SPT+{\it Planck} CMB lensing map and the methodology of the cosmological analysis. In \citep[][hereafter \citetalias{y3-nkgkmeasurement}]{y3-nkgkmeasurement}, we presented the measurements of the cross-correlation probes \nkgk{}, a series of diagnostic tests of the measurements, and cosmological constraints from this cross-correlation combination. In this paper (\textsc{Paper III}), we present the joint cosmological constraints from all the \sixtwo{} probes, and tests of consistency between various combinations of two-point functions.

The plan of the paper is as follows.  In \S\ref{sec:data} we describe the data sets from DES, SPT and {\it Planck} that we use in this analysis, and in \S\ref{sec:model} we provide an abridged summary of our model for the correlation function measurements. In \S\ref{sec:results}, we present cosmological constraints from the joint analysis of cross-correlations between DES and CMB lensing measurements from SPT and {\it Planck}, and discuss several tests of the robustness of these constraints enabled by the cross-correlation measurements. We conclude in \S\ref{sec:discussion}.

\section{Data from DES, SPT and Planck}
\label{sec:data}

\begin{figure}
	\includegraphics[width= \columnwidth]{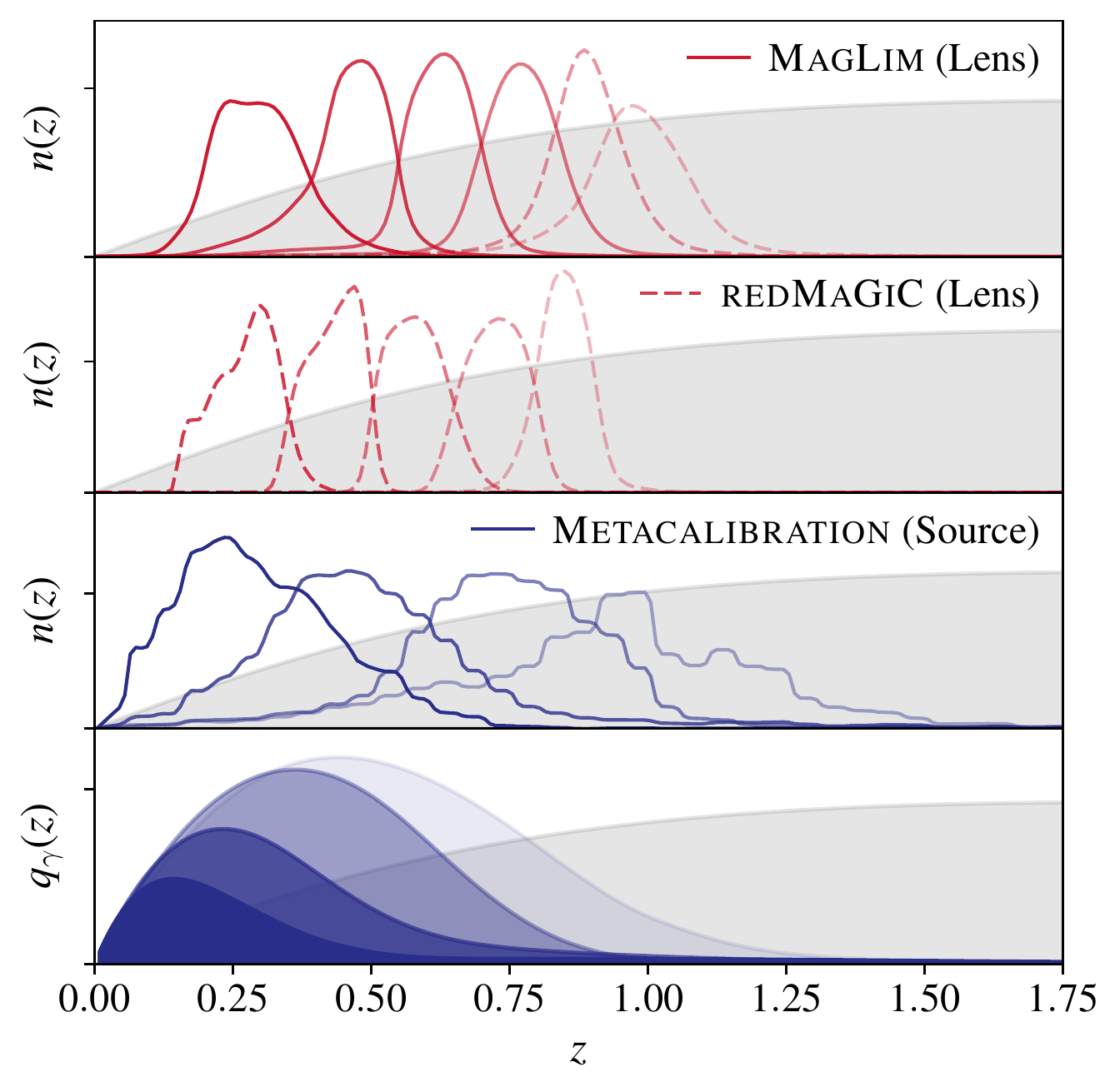}
    \caption{Redshift distributions of the galaxy samples considered in this work.  The \maglim{} (top panel) and \redmagic{} (second from top) lens galaxy samples  are used to measure the galaxy overdensity, while the \metacal{}  (third from top) source galaxy samples are used to measure weak lensing. Our main cosmological results use only the first four bins of the \maglim{} sample (solid lines). We perform tests with alternate samples (dashed lines) for exploratory and diagnostic purposes. In the bottom panel we show the lensing kernels (Eq.~\ref{eq:q_gamma}) corresponding to the source galaxies (blue).  The grey band in every panel represents the CMB lensing kernel (Eq.~\ref{eq:q_kappa}).}
    \label{fig:nofz}
\end{figure}

DES \citep{Flaugher2005} is a photometric survey in five broadband filters ($grizY$), with a footprint of nearly $5000 \; {\rm deg}^2$ of the southern sky, imaging hundreds of millions of galaxies. It employs the 570-megapixel Dark Energy Camera \citep[DECam,][]{Flaugher2015} on the Cerro Tololo Inter-American Observatory (CTIO) 4m Blanco telescope in Chile. We use data from the first three years (Y3) of DES observations. The foundation of the various DES Y3 data products is the Y3 Gold catalog described in \cite{y3-gold}, which achieves a depth of S/N$\sim$10 for extended objects up to i$\sim$23.0 over an unmasked area of $4143 \; {\rm deg}^2$. In this work, we consider two types of galaxy samples: {\it lens galaxies} that are used as biased tracers of the underlying density field, and {\it source galaxies} which are used to measure the shape-distorting effects of gravitational lensing. We use the same galaxy samples as in the DES \threetwo{} analysis \citep{y3-3x2ptkp}. That is, the lens galaxies are taken from the four-redshift bin \maglim{} sample described in \citep{y3-2x2maglimforecast}, and the source galaxy shapes are taken from the four-redshift bin \textsc{MetaCalibration} sample described in \citep{y3-shapecatalog}.  We will additionally consider lens galaxies from the \redmagic{} sample described in \citep{y3-2x2ptbiasmodelling}.  In particular, we will investigate the potential systematic biases that led to that sample not being used as the baseline cosmology sample in \cite{y3-3x2ptkp}.  The redshift distributions for the \maglim{}, \redmagic{}, \metacal{} samples are shown in Fig.~\ref{fig:nofz}. 

As mentioned above, we use two CMB lensing maps in this work: one covering the SPT-SZ footprint that uses data from SPT-SZ and \textit{Planck} (with an overlapping area of $\sim1800$ deg$^{2}$), and a second that covers the northern part of the DES survey that uses only \textit{Planck} data (with an overlapping area of $\sim2200$ deg$^{2}$).  Together, these two CMB lensing maps cover the full DES Y3 survey region.  Since the noise levels and beam sizes of SPT-SZ and \textit{Planck} are different, the resulting CMB lensing maps must be treated separately in our analysis.  In \citetalias{y3-nkgkmeasurement} we tested the consistency between the cosmological constraints from these two patches, finding good agreement.

\section{Modeling and Measurements}
\label{sec:model}

The theoretical framework we use in this analysis is laid out in \citetalias{y3-nkgkmethods} and \cite{y3-generalmethods}.  The full \sixtwo{} data vector consists of six two-point functions.  Since there is little correlation between \fivetwo{} and the CMB lensing autocorrelation measurements from {\it Planck}, we combine the corresponding constraints at the likelihood level; this approximation is validated in \citetalias{y3-nkgkmethods}.  

We fit the \sixtwo{} data to two different cosmological models: a spatially flat, cosmological constant and cold dark matter model, and a cosmological model where the equation of state parameter of dark energy, $w$, is additionally allowed to vary. Following the DES convention, we will refer to these models as $\Lambda$CDM and $w$CDM; note, though, that we allow the sum of the neutrino masses to vary in both of these analyses.

The modeling of the \fivetwo{} correlations begins with the auto and cross-power spectrum of the three fields ($\delta_{\rm g}$, $\gamma$, $\kappa_{\rm CMB}$).  For the correlation functions other than galaxy clustering, we use the Limber approximation \cite{Limber:1953}:
\begin{equation}
C^{X^{i} Y^{j}}(\ell) =\int d\chi \frac{q^{i}_{X}(\chi)q^j_{Y} (\chi)}{\chi^2} P_{\rm NL} \left( \frac{\ell+1/2}{\chi}, z(\chi)\right),
\label{eq:Cl_basic}
\end{equation}
where $X,Y \in \{\delta_g, \gamma, \kappa_{\rm CMB} \}$, $i,j$ labels the redshift bin, $P_{\rm NL}(k,z)$ is the non-linear matter power spectrum, which we compute using \textsc{CAMB} and \textsc{Halofit} \citep{camb,Takahashi:2012}, $\chi$ is the comoving distance from the observer, and $z(\chi)$ is the redshift corresponding to $\chi$. The weighting functions, $q(\chi)$, describe how the different probes respond to large-scale structure at different distances, and are given by
\begin{equation}
q^i_{\delta_g}(\chi) = b^i(k,z(\chi)) 
n_{\delta_{\rm g}}^i(z(\chi)) \frac{dz}{d\chi}
\label{eq:q_deltag}
\end{equation}
\begin{equation}
q^{i}_{\gamma}(\chi) = \frac{3H_0^2 \Omega_{\rm m}}{2c^2} \frac{\chi}{a(\chi)} \int_{\chi}^{\chi_h} d\chi' 
n_{\gamma}^i(z(\chi')) 
\frac{dz}{d\chi'}
\frac{\chi' -\chi}{\chi'}, 
\label{eq:q_gamma}
\end{equation}
\begin{equation}
q_{\kappa_{\rm CMB}}(\chi) = \frac{3H_0^2 \Omega_{\rm m}}{2c^2} \frac{\chi}{a(\chi)} \frac{\chi^{*} -\chi}{\chi^{*}}, 
\label{eq:q_kappa}
\end{equation}
where $H_0$ and $\Omega_{\rm m}$ are the Hubble constant and matter density parameters, respectively, $a(\chi)$ is the scale factor corresponding to comoving distance $\chi$, $b(k,z)$ is galaxy bias as a function of scale ($k$) and redshift, $n^i_{{\delta_{\rm g}}/\gamma}(z)$ are the normalized redshift distributions of the lens/source galaxies in bin $i$. $\chi^*$ denotes the comoving distance to the CMB last scattering surface.  The model for galaxy clustering is computed without using the Limber approximation, as described in \cite{y3-generalmethods}.  The angular-space correlation functions are then computed from the auto- and cross-spectra as described in  \cite{y3-generalmethods,y3-nkgkmethods}. 

In addition to the basic modeling, described above, we also consider several other physical and observational effects.  We list these below but refer the readers to \citetalias{y3-nkgkmethods} and \cite{y3-generalmethods} for details. 
\begin{itemize}
    \item {\bf Galaxy bias:} Our baseline model assumes linear galaxy bias, but we also explore the potential improvement from using a nonlinear galaxy bias model and including smaller angular scales in our analysis, as described in \citep{y3-2x2ptbiasmodelling} and \citetalias{y3-nkgkmethods}.
    \item {\bf Intrinsic alignments (IA):} We use the Tidal Alignment and Tidal Torquing \citep[TATT,][]{blazek2019} model to describe the effect of galaxy intrinsic alignments.  We consider an alternate IA model in Appendix~\ref{app:photoz_IA}.
    \item {\bf Lens magnification:} Gravitational lensing by foreground mass changes the observed projected number density of lens galaxies as a result of geometric dilution and modulation of galaxy flux and size. We model this effect based on measurements in simulations as described in \citep{y3-2x2ptmagnification,y3-generalmethods}. 
    \item {\bf Redshift uncertainties:} There are uncertainties associated with the estimation of the redshift distributions of the lens and the source sample, which we model as described in \citep{y3-sompz, y3-sourcewz, y3-lenswz}.  In \citep{y3-hyperrank}, an alternate approach to marginalizing over uncertainties in the redshift distributions was also considered, which we explore in Appendix~\ref{app:photoz_IA}. 
    \item {\bf Shear calibration uncertainties:} We include a prescription for uncertainties in shear calibration as described in \cite{y3-3x2ptkp}.  We estimate uncertainties in the shear measurements using realistic image simulations as described in \citep{y3-imagesims}.
    \item {\bf CMB map filtering:} In order to suppress very small-scale noise in the CMB lensing cross-correlations, we apply filtering to the CMB lensing maps.  This filtering is included in the model as described in \citetalias{y3-nkgkmethods}.
    \item {\bf Point mass marginalization:} The correlation functions at small scales are impacted by baryonic effects that are challenging to model, such as galaxy formation.   This is particularly problematic for \galshear{}: changes in e.g. the masses of the lens galaxies at very small scales can impact the large-scale \galshear{} because tangential shear is a non-local quantity.  To reduce sensitivity of our analysis to small-scale effects in \galshear{}, we therefore adopt the point mass marginalization approach of \citep{MacCrann:2020}, which involves modifying the covariance matrix of \galshear{}.  
\end{itemize}

We measure the two-point angular correlation functions of the data using the fast tree-based algorithm \textsc{TreeCorr} \cite{Jarvis2004} as described in \citep{y3-galaxyclustering,y3-gglensing,y3-cosmicshear1,y3-cosmicshear2,y3-nkgkmeasurement}.  The shear measurements define a spin-2 field on the sky, and there are several ways of decomposing this field for the purposes of measuring two-point functions.  For measuring \shearshear{}, we use the $\xi_{+}$ and $\xi_{-}$ decomposition, while for measuring \galshear{}, we consider the correlation only with tangential shear, $\gamma_{\rm t}$ \citep{Bartelmann2001}.  The covariance matrix associated with these measurements is constructed by combining an analytical halo model covariance, analytical lognormal covariance, and empirical noise estimation from simulations \citep{y3-covariances,y3-nkgkmethods}.

For the final parameter inference, we assume a Gaussian likelihood.\footnote{See e.g. \citep{Lin2020} for tests of the validity of this assumption in the context of cosmic shear, which would also apply here.} The priors imposed on the model parameters are shown in Table~\ref{table:prior} in Appendix~\ref{app:priors}. The modeling and likelihood framework is built within the \textsc{CosmoSIS} package \cite{cosmosis}. We generate parameter samples using the nested sampler \textsc{PolyChord} \cite{polychord}.

Due to uncertainties in the modeling of the correlation functions on small scales (e.g., nonlinear galaxy bias and baryonic effects on the matter power spectrum), in our likelihood analysis we remove the small-scale measurements that could potentially bias our cosmological constraints. The procedure of determining these ``scale cuts'' is described in \citep{y3-generalmethods} and \citetalias{y3-nkgkmethods}.  Note that the choice of angular scales used in the analysis varies somewhat depending on whether we assume a linear or nonlinear galaxy bias model.  We focus on the results with linear bias, but consider the results from the nonlinear bias analysis in Appendix~\ref{app:nonlinear_bias}.

In each of the cosmological analyses performed in this work, we include a separate likelihood constructed using a set of ratios of galaxy-galaxy lensing measurements on small scales \citep{y3-shearratio}. These lensing ratios are found to primarily constrain parameters describing the intrinsic alignment model and redshift biases, and are effectively independent of the \fivetwo{} data vector.

We utilize two different statistical metrics to assess the consistency of the DES and CMB-lensing cross-correlation measurements, both internally (i.e. between the different two point functions that we measure) and with other cosmological probes.  To assess internal consistency, we primarily rely on the posterior predictive distribution (PPD) methods described in \cite{y3-inttensions}.  For these assessments, we will quote $p$-values, with $p < 0.01$ taken as significant evidence of inconsistency.  To assess external consistency, we rely on the parameter difference methods developed in \cite{Raveri:2021}.   For this metric, we will quote differences between parameter constraints in terms of effective $\sigma$ values, corresponding to the probability values obtained from the non-Gaussian parameter difference metric.  When computing the goodness of fit of our measurements to a particular model, we again rely on the PPD methodology, as discussed in \cite{y3-inttensions}.  In this case, the associated $p$-values can be thought of as a generalization of the classical $p$-value computed from the $\chi^2$ statistic that correctly marginalizes over parameter uncertainty.

\section{Cosmological constraints}
\label{sec:results}

\begin{table*}[]
    \centering
    \begin{tabular}{|c|ccc|c|c|c|}
    \hline
        Probe & $\sigma_{8}$ & $\Omega_{\rm m}$ & $S_8$ & GoF $p$-value & Comments \\ \hline
        \threetwo{} & $0.733^{+0.039}_{-0.049}$ & $0.339^{+0.032}_{-0.031}$ & $0.776\pm 0.017$ & 0.023 & \cite{y3-3x2ptkp}\\
        \nkgk{} & $0.78\pm 0.07$ & $0.27^{+0.03}_{-0.05}$ & $0.74\pm 0.03$ & 0.50 & CMB lensing cross-correlations, \citetalias{y3-nkgkmeasurement}\\ 
        \galshear+\nk+\gk & $0.768\pm 0.071$ & $0.303^{+0.036}_{-0.059}$ & $0.765\pm 0.025$ & 0.063 & All cross-correlations, \S\ref{sec:xcorrs} \\
        \fivetwo{} & $0.724^{+0.038}_{-0.043}$ & $0.344\pm 0.030$ & $0.773\pm 0.016$& 0.062 & \S\ref{sec:baseline} \\
        \sixtwo{} & $0.785\pm 0.029$ & $0.306\pm 0.018$ & $0.792\pm 0.012$ & --- & \S\ref{sec:baseline}  \\
        \hline
    \end{tabular}
    \caption{$\Lambda$CDM constraints on $\Omega_{\rm m}$, $\sigma_{8}$ and $S_{8}\equiv \sigma_8 (\Omega_{\rm m}/0.3)^{0.5}$ using different subsets of the \sixtwo{} two-point functions. The $p$-values correspond to the goodness of fit, as calculated using the PPD methodology. All results here use 4-bin \maglim{} lens sample and linear galaxy bias.  For the \sixtwo{} combination, we do not quote a goodness of fit because the CMB lensing autospectrum is treated as an external probe.  Rather, we use the parameter difference metric to assess tension between \fivetwo{} and \kk{} (see \S\ref{sec:model}). }
    \label{tab:param_fiducial}
\end{table*}

\subsection{Baseline cosmological constraints}
\label{sec:baseline}

\subsubsection{$\Lambda$CDM}

We first present constraints on $\Lambda$CDM from the joint analysis of two-point functions involving DES galaxy position and lensing measurements, and measurements of CMB lensing from SPT and {\it Planck}.   Following \cite{y3-3x2ptkp}, all of the results in this subsection use the four redshift bin \maglim{} lens galaxy sample.

\begin{figure}
	\includegraphics[width=0.99 \columnwidth]{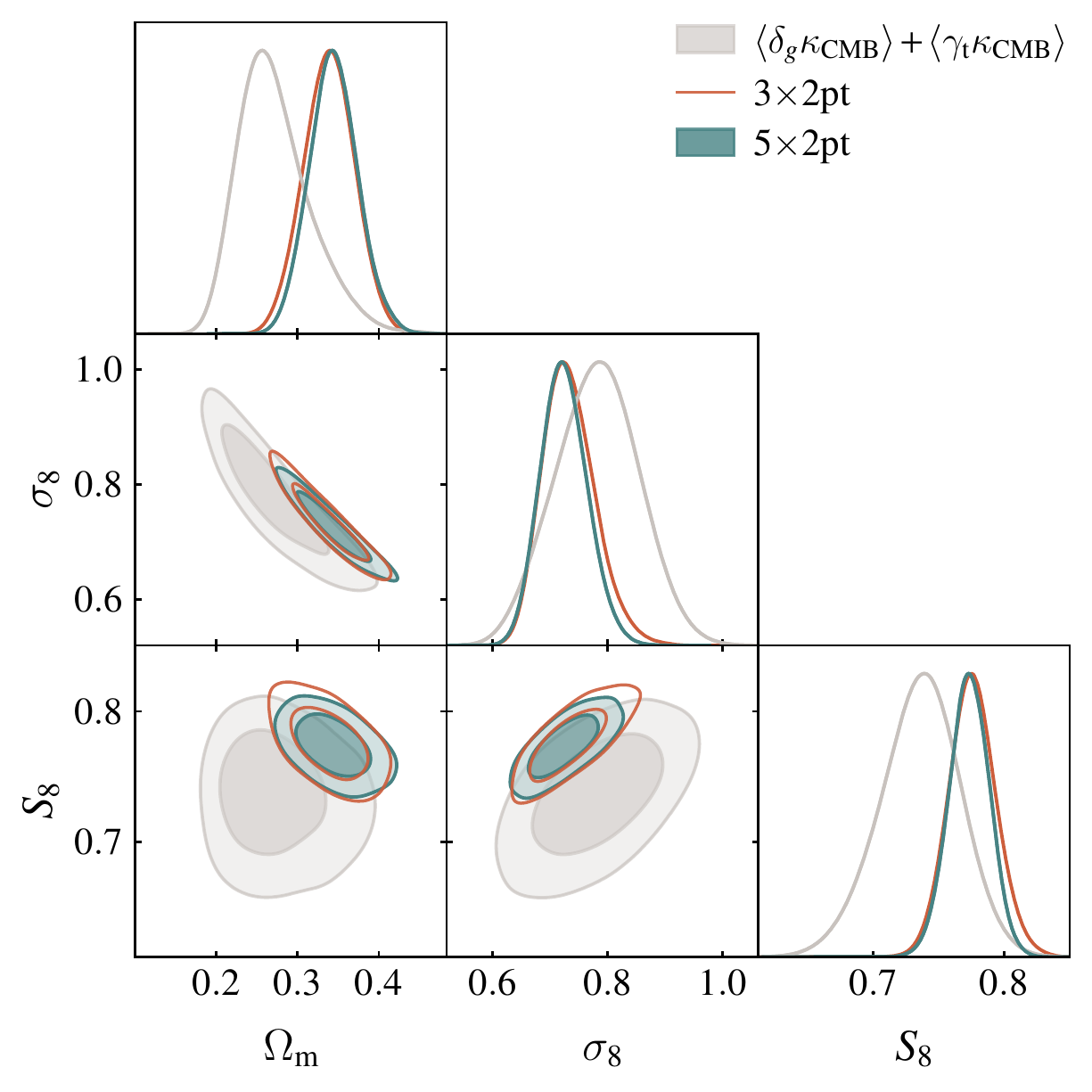}
    \caption{$\Lambda$CDM constraints from the DES Y3 \threetwo{} measurements (red), cross-correlations between DES Y3 galaxies and shears with SPT+{\it Planck} CMB lensing (grey), and from the joint analysis of all five two-point functions (teal).  The constraints from \threetwo{} are in acceptable agreement with the CMB lensing cross-correlations, justifying the joint analysis of \fivetwo{}.}
    \label{fig:nkgk_3x2_5x2}
\end{figure}

Fig.~\ref{fig:nkgk_3x2_5x2} shows how the constraints from the CMB lensing cross-correlations \nkgk{} compare to those from \threetwo{}. The resulting 68\% credible intervals on $\sigma_{8}$, $S_{8}$, and $\Omega_{\rm m}$ computed from the marginalized \threetwo{} and \nk{}+\gk{} posteriors are summarized in Table~\ref{tab:param_fiducial}.  In the same table, we list the goodness of fit $p$-values for \threetwo{} and \nk{} + \gk{}, computed using the PPD formalism.  As noted in \cite{y3-3x2ptkp}, the goodness of fit for \threetwo{} alone is not particularly high, but is still above our threshold of $p=0.01$.  The goodness of fit for \nk{} + \gk{} is acceptable, as described in \cite{y3-nkgkmeasurement}.  While the cross-correlations prefer somewhat lower $\Omega_{\rm m}$ and higher $\sigma_8$, they are statistically consistent with \threetwo{}.  Using the PPD formalism, we find $p=0.347$ when comparing the two data subsets, indicating acceptable consistency.  We are therefore justified in combining the constraints to form \fivetwo{}, shown with the teal contours in the figure.  

Given the weaker constraining power of \nkgk{} relative to \threetwo{}, the \fivetwo{} constraints are not much tighter than the \threetwo{} constraints: we find an improvement of roughly 10\% in the precision of the marginalized constraints on $\Omega_m$ and $S_8$ (see Table~\ref{tab:param_fiducial}).  The goodness of fit for the full \fivetwo{} data vector is $p=0.062$, indicating an acceptable fit.   

\begin{figure*}
	\includegraphics[width=0.6 \textwidth]{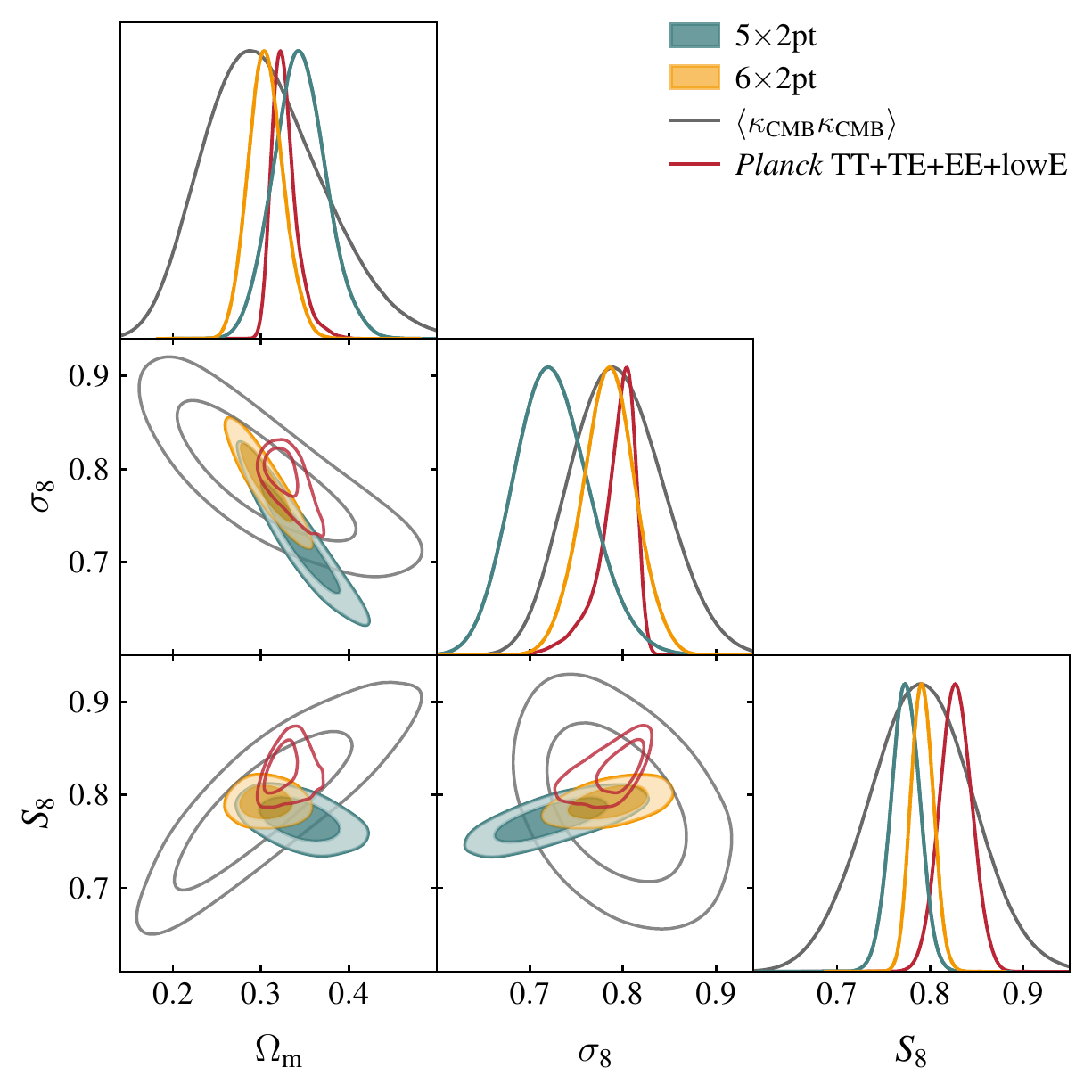}
    \caption{$\Lambda$CDM constraints from our \fivetwo{} analysis (teal) are compared to those from the {\it Planck} CMB lensing autospectrum measurements (grey).  The two are in acceptable agreement, justifying the joint analysis of \sixtwo{} (orange), which yields significantly tighter constraints due to degeneracy breaking.  Also shown are parameter constraints from \planck{} measurements of primary CMB fluctuations (TT+TE+EE+lowE, dark red).
    }
    \label{fig:5x2_kk_6x2}
\end{figure*}

In Fig.~\ref{fig:5x2_kk_6x2} we compare the constraints from \fivetwo{} with those from the CMB lensing autospectrum \kapkap{}{}.  Owing to the high redshift of the CMB source plane, the CMB lensing-only contour has a different degeneracy direction than \fivetwo{}, resulting in a weaker constraint when projecting to the $\Omega_{\rm m}$ direction, but a comparable constraint in the $\sigma_8$ direction.  While the CMB lensing autospectrum prefers somewhat higher $\sigma_8$ than \fivetwo{}, the constraints are generally consistent.  Because the CMB lensing autospectrum measurements are treated as an independent probe, we quantify the tension between these measurements and \fivetwo{} using the parameter shift metric, finding a difference of $0.8\sigma$, indicating no evidence of significant tension.  We therefore combine the two to generate constraints from all six two-point functions, \sixtwo{}, shown with the orange contour in the figure. Due to degeneracy breaking, the joint analysis leads to notably tighter constraints on both $\Omega_{\rm m}$ and $\sigma_8$.  The 1D posterior constraints on these parameters from \sixtwo{} are summarized in Table~\ref{tab:param_fiducial}.  Fig.~\ref{fig:5x2_kk_6x2} also shows constraints from {\it Planck} measurements of CMB temperature and polarization fluctuations \cite{Planck:cosmo}.  We will assess consistency between our measurements and the {\it Planck} measurements in \S\ref{sec:tension_planck}.

\begin{figure}
	\includegraphics[width=0.99 \columnwidth]{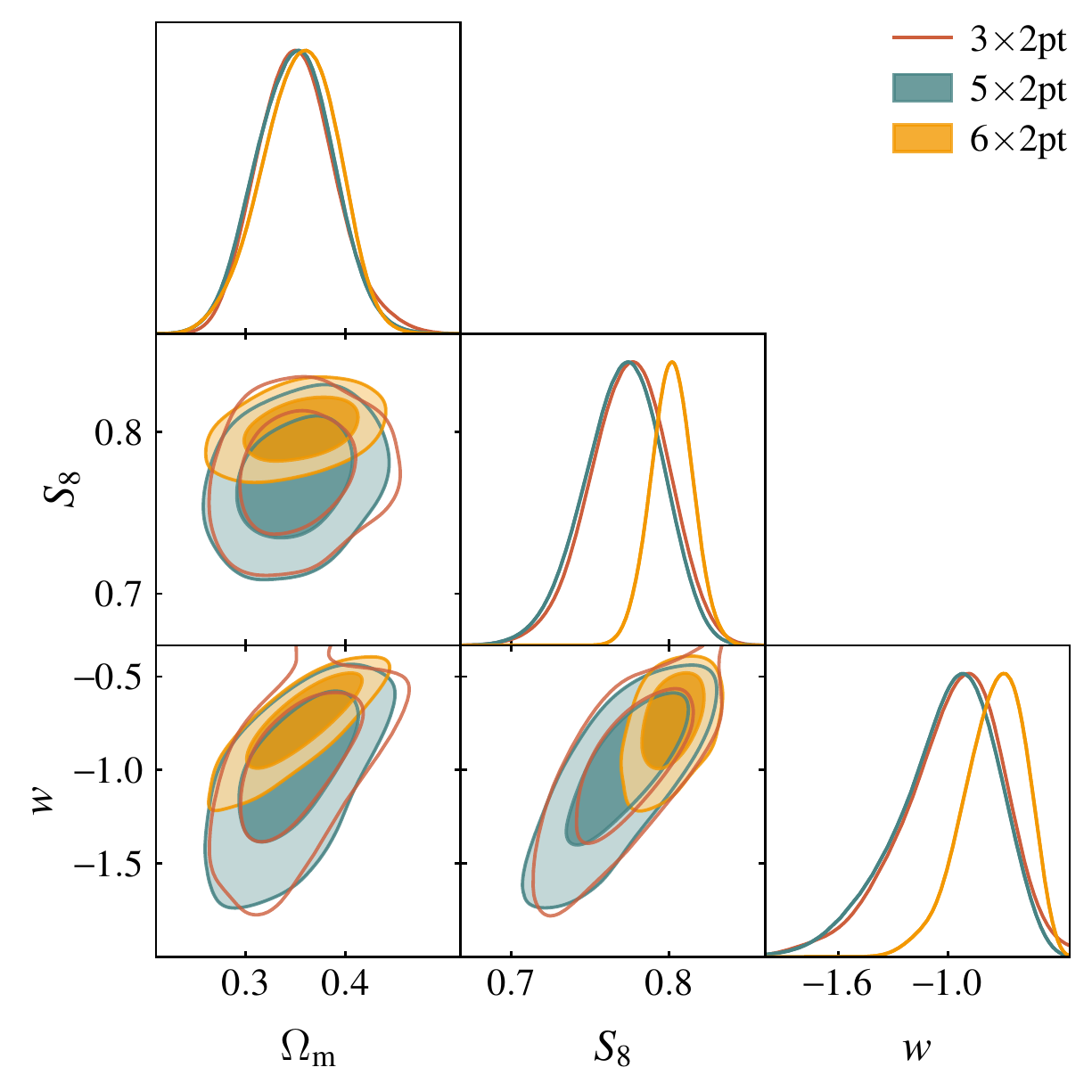}
    \caption{Constraints on $w$CDM from different combinations of two-point functions.  The \fivetwo{} constraints (teal) on this model are essentially identical to those of \threetwo{} (red).  Adding the CMB lensing autospectrum information in the joint \sixtwo{} analysis (orange) significantly improves the parameter constraints on $w$CDM.
    }
    \label{fig:maglim_wcdm}
\end{figure}

\subsubsection{$w$CDM}

We now consider constraints on $w$CDM, the cosmological model with a constant equation-of-state parameter of dark energy, $w$.  The constraints from \threetwo{}, \fivetwo{} and \sixtwo{} are shown in Fig.~\ref{fig:maglim_wcdm}.  We find that there is little improvement in constraining power on $w$CDM when adding the CMB lensing cross-correlations to \threetwo{}.  Adding the \kk{} correlation, however, significantly impacts the constraints, presumably because this correlation function adds additional information about structure at $z \gtrsim 1$.  The \sixtwo{} analysis yields $w = -0.75^{+0.20}_{-0.14}$, $S_8 = 0.801\pm 0.013$,  and $\Omega_{\rm m} = 0.354^{+0.041}_{-0.035}$.  Therefore, the constraints on the dark-energy equation of state parameter are largely consistent with the cosmological-constant scenario of $w=-1$, and the constraints on $\Omega_{\rm m}$ and $S_8$ are consistent with those obtained assuming $\Lambda$CDM.

\subsection{Robustness tests}

In addition to improving cosmological constraints relative to the DES-only \threetwo{} analysis, a significant motivation for cross-correlating DES with CMB lensing is to test the robustness of the DES-only constraints.  The cross-correlations probe the same large-scale structure as the DES \threetwo{} analysis, but with sensitivity to different potential sources of systematic bias, making them powerful cross-checks on the DES results.  In this section, we subject the \sixtwo{} data vector to several tests of internal consistency.  

\begin{figure}
	\includegraphics[width=1.00 \columnwidth]{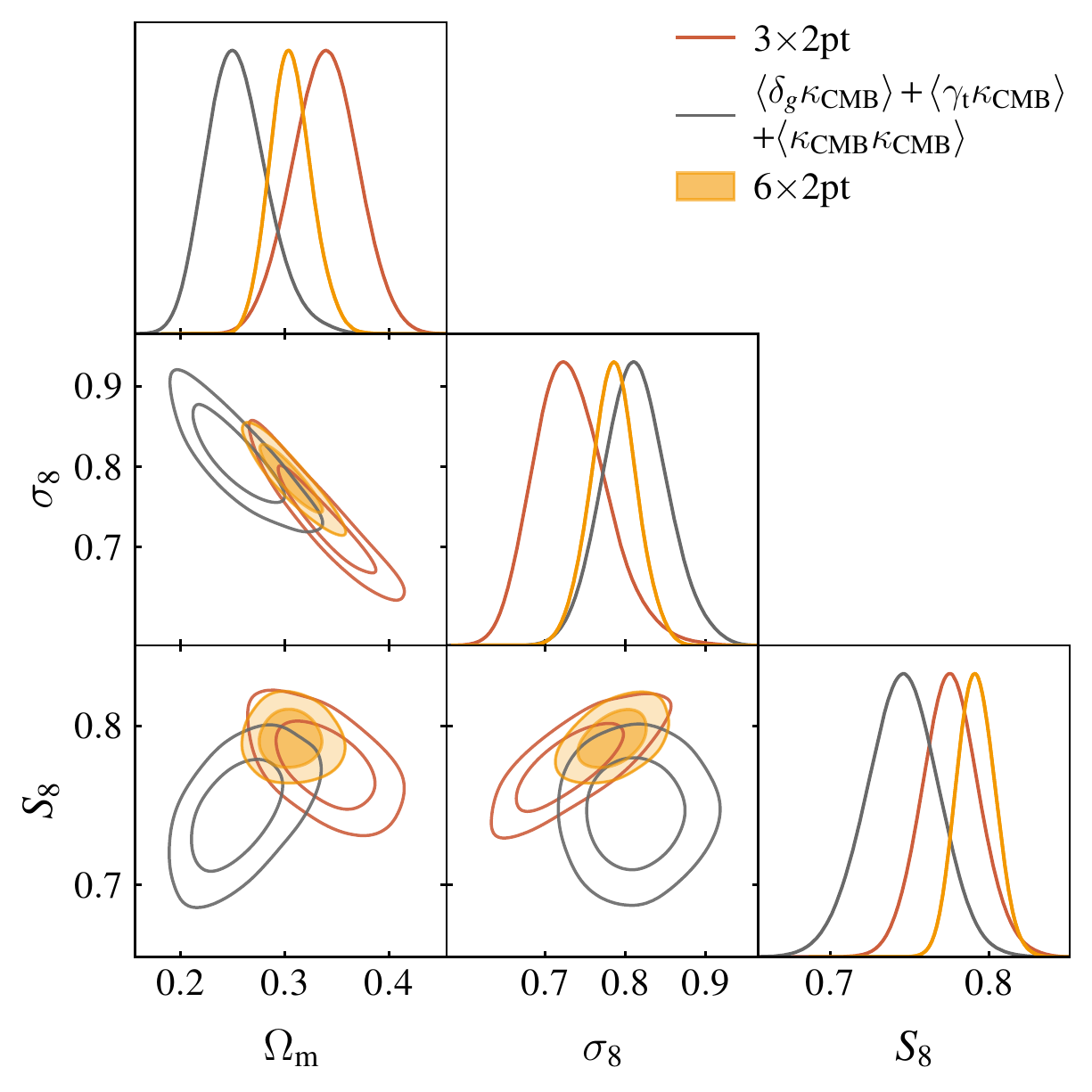}
    \caption{Comparison of constraints on $\Lambda$CDM from \threetwo{} (red) with the constraints from the other probes of \sixtwo{}, i.e. \nk+\gk +\kk (grey).  The joint analysis of both (\sixtwo{}) is shown in orange.  The two subsets of the full \sixtwo{} analysis are in reasonable agreement.  The \sixtwo{} analysis prefers higher $S_8$ than either of the two subsets.}
    \label{fig:other_3x2}
\end{figure} 

\subsubsection{\threetwo{} vs. \otherthreetwo{}}
\label{sec:other_3x2}

We first assess the internal consistency of the \sixtwo{} combination of probes by comparing constraints from \threetwo{}  to the other three two-point functions making up \sixtwo{}, which we call \otherthreetwo{} (i.e. \nk{}+\gk{}+\kk{}).  This comparison is shown in Fig.~\ref{fig:other_3x2}.  We find that the constraining power from \otherthreetwo{} is very similar to that of \threetwo{}.  Because \otherthreetwo{} does not constrain galaxy bias or intrinsic alignment parameters very well, applying the PPD methodology to test consistency between \threetwo{} and \otherthreetwo{} is not well motivated.  However, we note that we have already tested the consistency of \threetwo{} with \nk{} + \gk{} (i.e. part of \otherthreetwo{}), finding acceptable agreement ($p = 0.347$). 

Fig.~\ref{fig:other_3x2} makes it clear why \sixtwo{} prefers a somewhat higher value of $S_8$ than \threetwo{}.  It is \textit{not} the case that \otherthreetwo{} prefers a higher value of $S_8$ than \threetwo{}; indeed, the opposite is true.  Rather, the slightly high value of $S_8$ found for \sixtwo{} is caused by the fact that \threetwo{} and \otherthreetwo{} have somewhat different degeneracy directions, and intersect at a high value of $S_8$ for both probes. 

\subsubsection{Cross-correlations}
\label{sec:xcorrs}

\begin{figure}
	\includegraphics[width=0.99\linewidth]{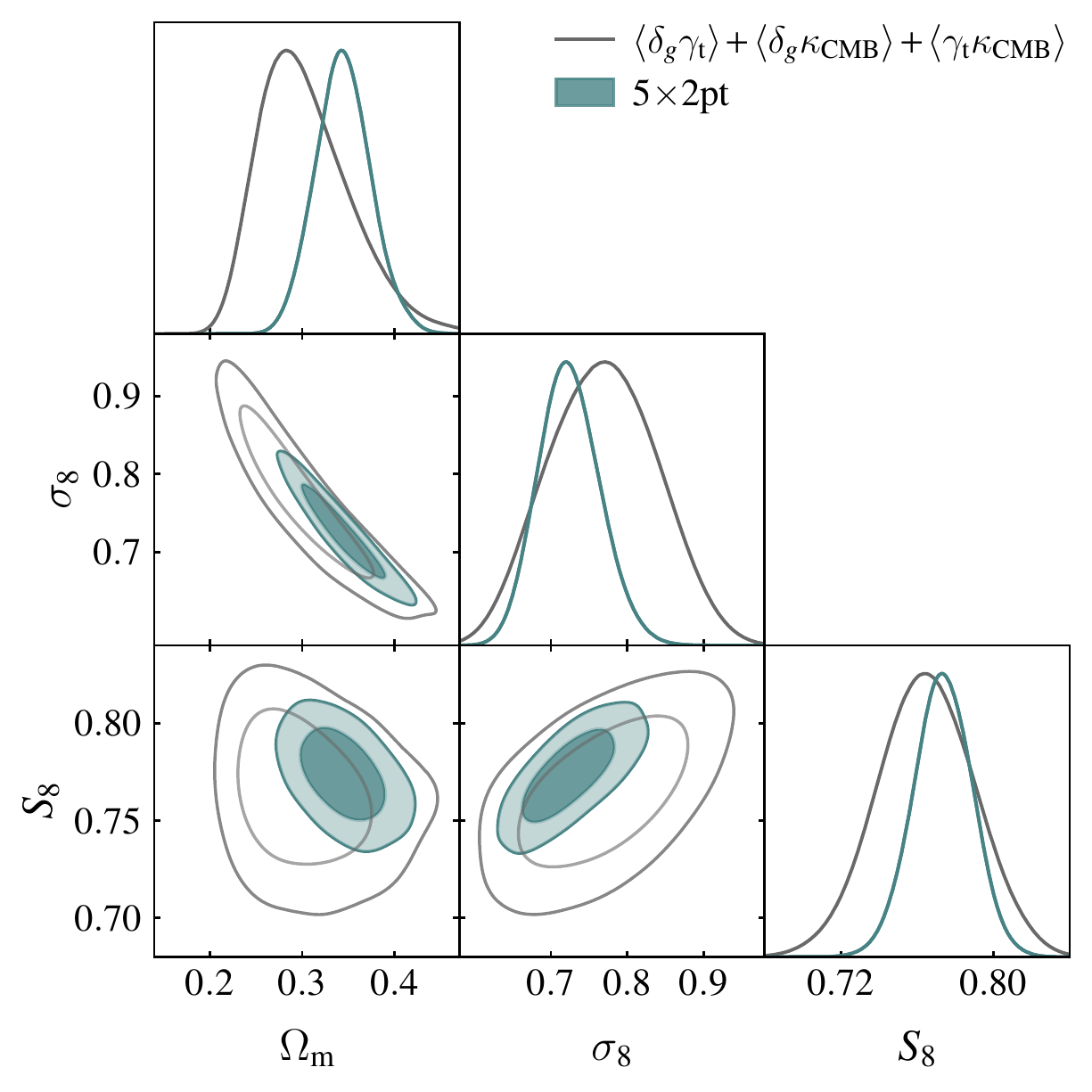}
    \caption{Comparison of constraints on $\Lambda$CDM resulting from \fivetwo{} (teal) to those that result from only cross-correlations between $\delta_g$, $\gamma$ and $\kappa_{\rm CMB}$ (grey).  Cross-correlations are expected to be robust to additive systematics that impact only a single field.  While some constraining power is lost by removing the auto-correlations, the resulting constraints on $S_8$ are  consistent with those of the baseline analysis, providing a powerful robustness test. }
    \label{fig:xcorrs_vs_5x2}
\end{figure}

Cross-correlations between different observables are generally expected to be more robust to systematic biases than auto-correlations of those observables.  Additive systematics that impact a single observable are expected to drop out of cross-correlations with another observable that has uncorrelated systematics.  In Fig.~\ref{fig:xcorrs_vs_5x2} we compare the cosmological constraints obtained  from \textit{only} cross-correlations to those from the full \fivetwo{}.  It is clear that removing the information from the auto-correlations --- particularly cosmic shear --- degrades the constraints somewhat.  However, we find that the value of $S_8$ inferred only from cross-correlations is consistent with that inferred from the full \fivetwo{} analysis.  This suggests that additive biases are unlikely to be having a major impact on the DES \threetwo{} cosmology results.   Using the PPD formalism to evaluate the goodness of fit of the cross-correlations conditioned on the posterior from \fivetwo{}, we find $p=0.054$, indicating an acceptable level of consistency between the \fivetwo{} constraints and the cross-correlations measurements.

\subsubsection{Lensing only}
\label{sec:lensing_only}

\begin{figure}
	\includegraphics[width=0.99\linewidth]{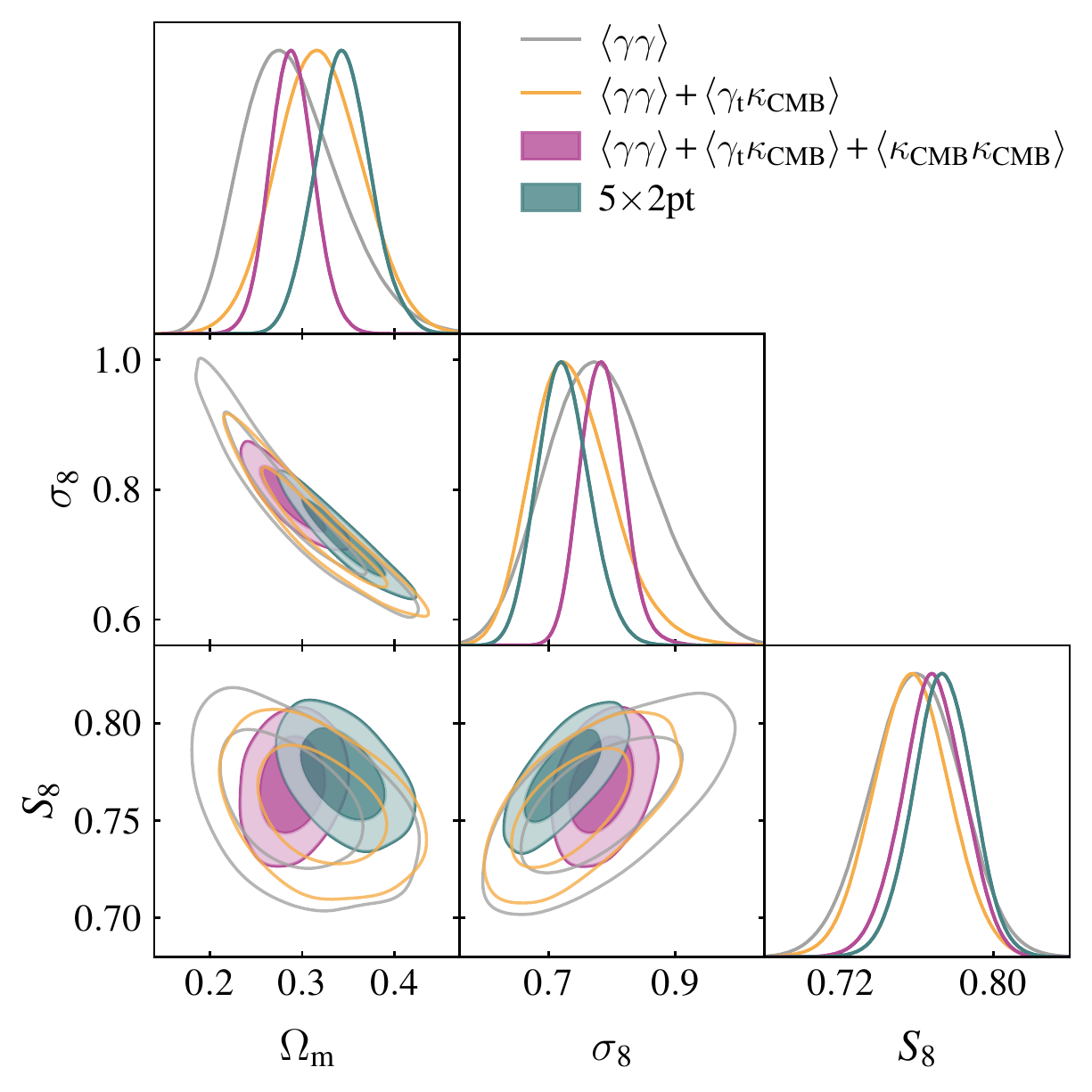}
    \caption{Comparison of baseline \fivetwo{} constraints on $\Lambda$CDM (teal) to constraints from various combinations of probes that only involve gravitational lensing.  The lensing-only constraints are consistent with our baseline result, suggesting that any systematics which might be impacting the galaxy overdensity measurements are not dramatically biasing our cosmological constraints.   }    \label{fig:lensing_only}
\end{figure} 

The relationship between galaxy overdensity and the underlying matter field --- galaxy bias --- presents a significant challenge for analyses of the galaxy distribution.  The baseline \threetwo{} results presented in \cite{y3-3x2ptkp} and the baseline cross-correlation results presented here assume a linear galaxy bias relation when modeling the galaxy field.  This model is known to break down at small scales, as investigated for the DES galaxy samples in \citep{y3-2x2ptbiasmodelling}.  More complex bias models, such as the perturbation theory-motivated model developed in \cite{Pandey:3dbias}, are also expected to have a limited range of validity. There is therefore value in performing analyses that use \textit{only} lensing information.

Another motivation to consider lensing-only analyses is that the DES galaxy overdensity measurements made with the \redmagic{} and  high-redshift \maglim{} galaxies show evidence of systematic biases (i.e. the samples shown with dashed lines in Fig.~\ref{fig:nofz}).   Measurements of galaxy-galaxy lensing with the \redmagic{} galaxies were shown to be inconsistent with clustering measurements using those galaxies \cite{y3-2x2ptbiasmodelling}.  This inconsistency suggests a potential problem with the \redmagic{} overdensity measurements, although it is not clear whether such issues could be impacting the galaxy-galaxy lensing measurements, clustering measurements, or both.  Similarly, galaxy-galaxy lensing and clustering measurements with the high-redshift \maglim{} galaxies were also found to be mutually inconsistent, contributing to a very poor goodness of fit to any of the cosmological models considered.  For these reasons, the high-redshift \maglim{} galaxies were removed from the cosmological analysis in \cite{y3-3x2ptkp}.  These issues, which we investigate further in \S\ref{sec:xlens}, further motivate a cosmological analysis that does not rely on galaxy overdensity measurements.

In Fig.~\ref{fig:lensing_only}, we present cosmological constraints from gravitational lensing only, namely the two-point functions of galaxy lensing and CMB lensing, and their cross-correlation.  The lensing-only analysis obtains cosmological constraints that are of comparable precision to those from the full \fivetwo{} analysis.  We find that the lensing-only analysis yields a constraint on $S_8$ that is in excellent agreement with the baseline analysis. 

\subsubsection{No galaxy lensing}
\label{sec:no_lensing}

\begin{figure}
	\includegraphics[width=0.99\linewidth]{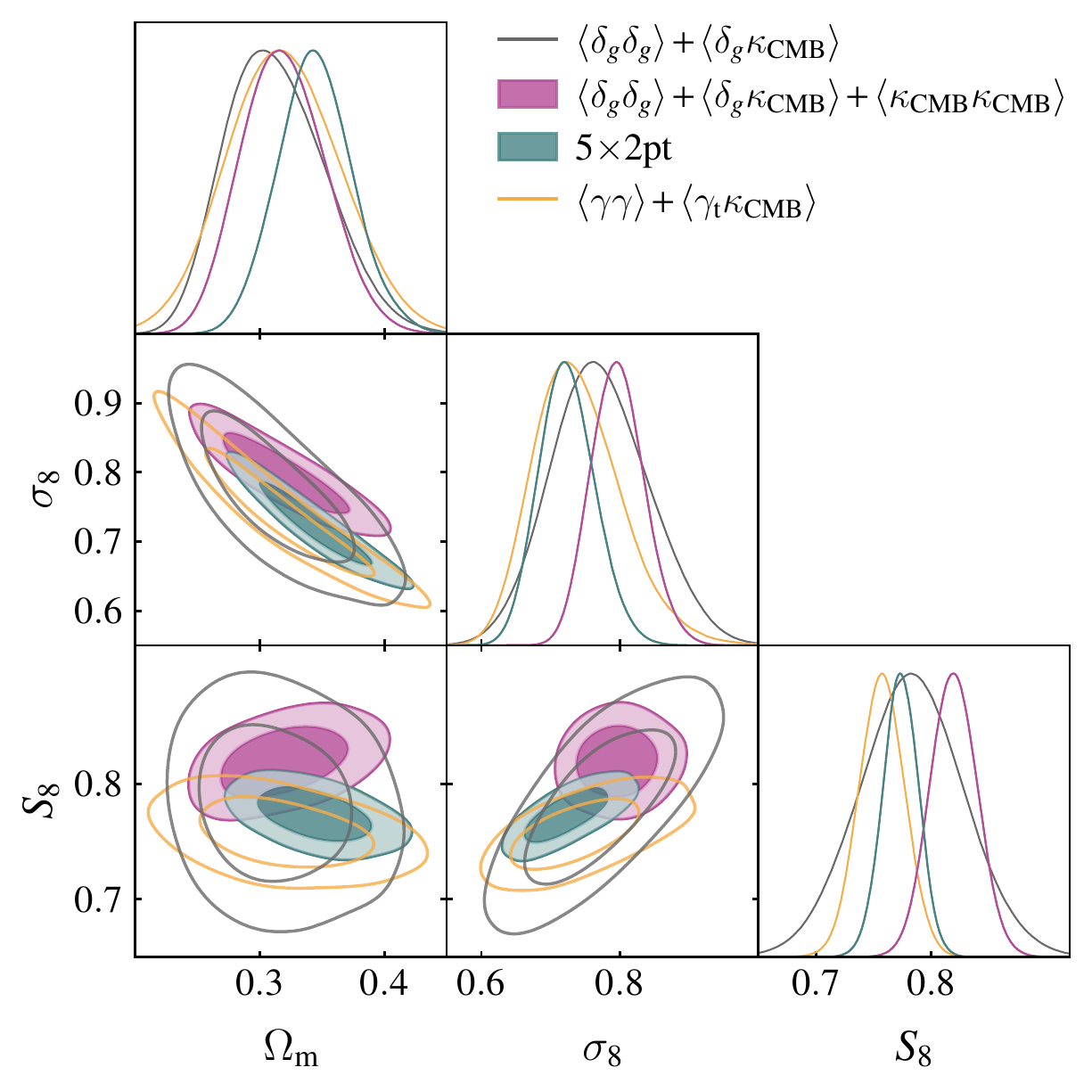}
    \caption{Comparison of baseline \fivetwo{} constraints on $\Lambda$CDM (teal) to constraints from those combinations of probes that do not rely on galaxy lensing (grey and purple). For reference, we also show the lensing-only constraints --- excluding the \kk{}, which is sensitive to higher redshifts --- with the orange curve. }
    \label{fig:nogal_lensing}
\end{figure} 

We also consider the constraints that result from those probes that do \textit{not} involve galaxy lensing.  The galaxy lensing measurements could in principle be biased by systematic errors in photometric redshifts of the source galaxies, shear calibration, or an incorrect intrinsic alignment model.  Such issues could bias constraints involving galaxy lensing, but would not impact the galaxy overdensity or CMB lensing measurements.  Fig.~\ref{fig:nogal_lensing} shows the constraints that result only from probes that do not include galaxy lensing (i.e. $\delta_g$ and $\kappa_{\rm CMB}$).   Again, we find that the results are  consistent with those of \fivetwo{}.  Fig.~\ref{fig:nogal_lensing} also shows the \shearshear{}+\gk{} constraints for comparison (i.e. lensing only, but excluding \kk{}, which receives contributions from higher redshifts than the other two-point functions).  We find that the constraints involving lens galaxy overdensities are consistent with the lensing-only constraints.  

\begin{figure*}
	\includegraphics[width=0.49\linewidth]{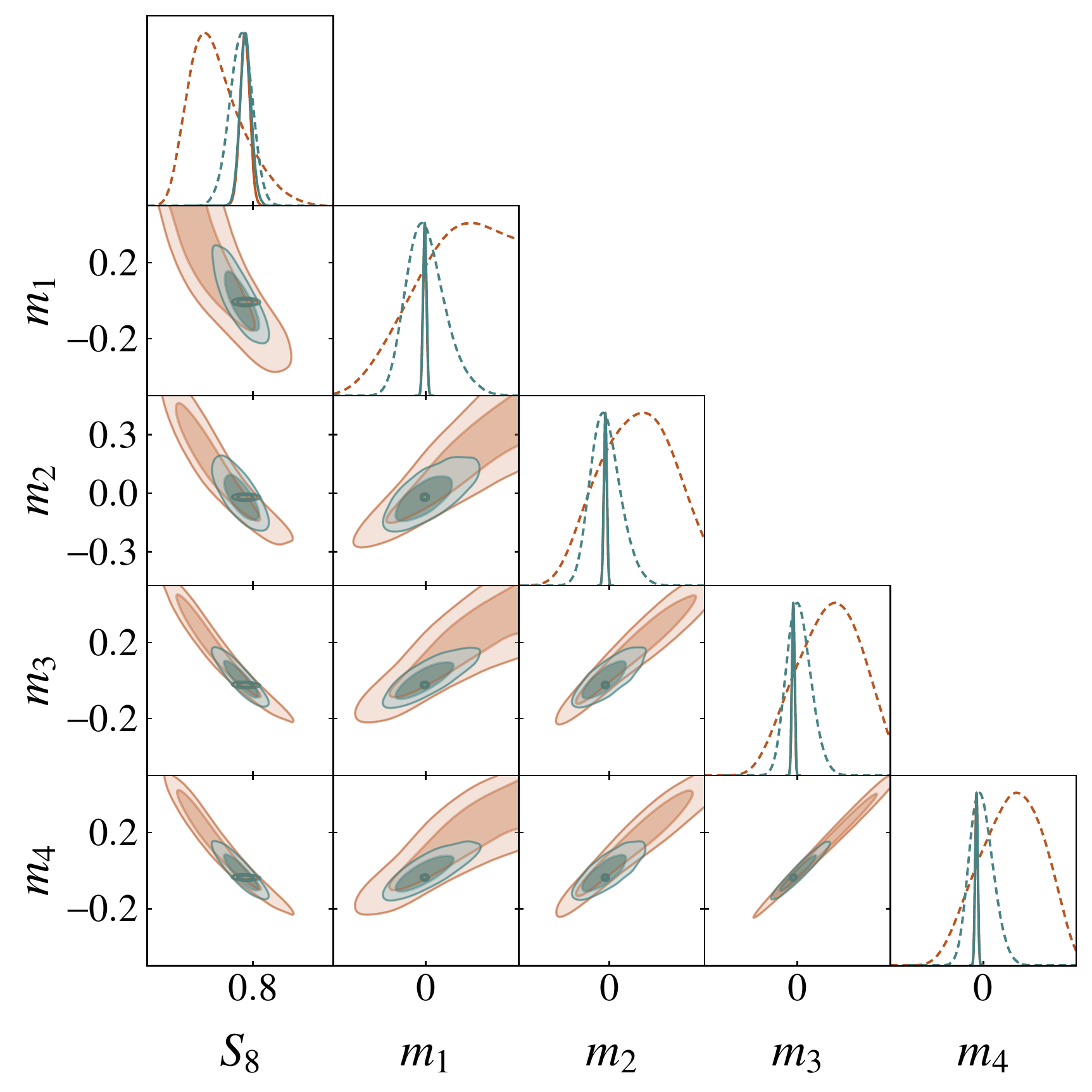}
	\includegraphics[width=0.49\linewidth]{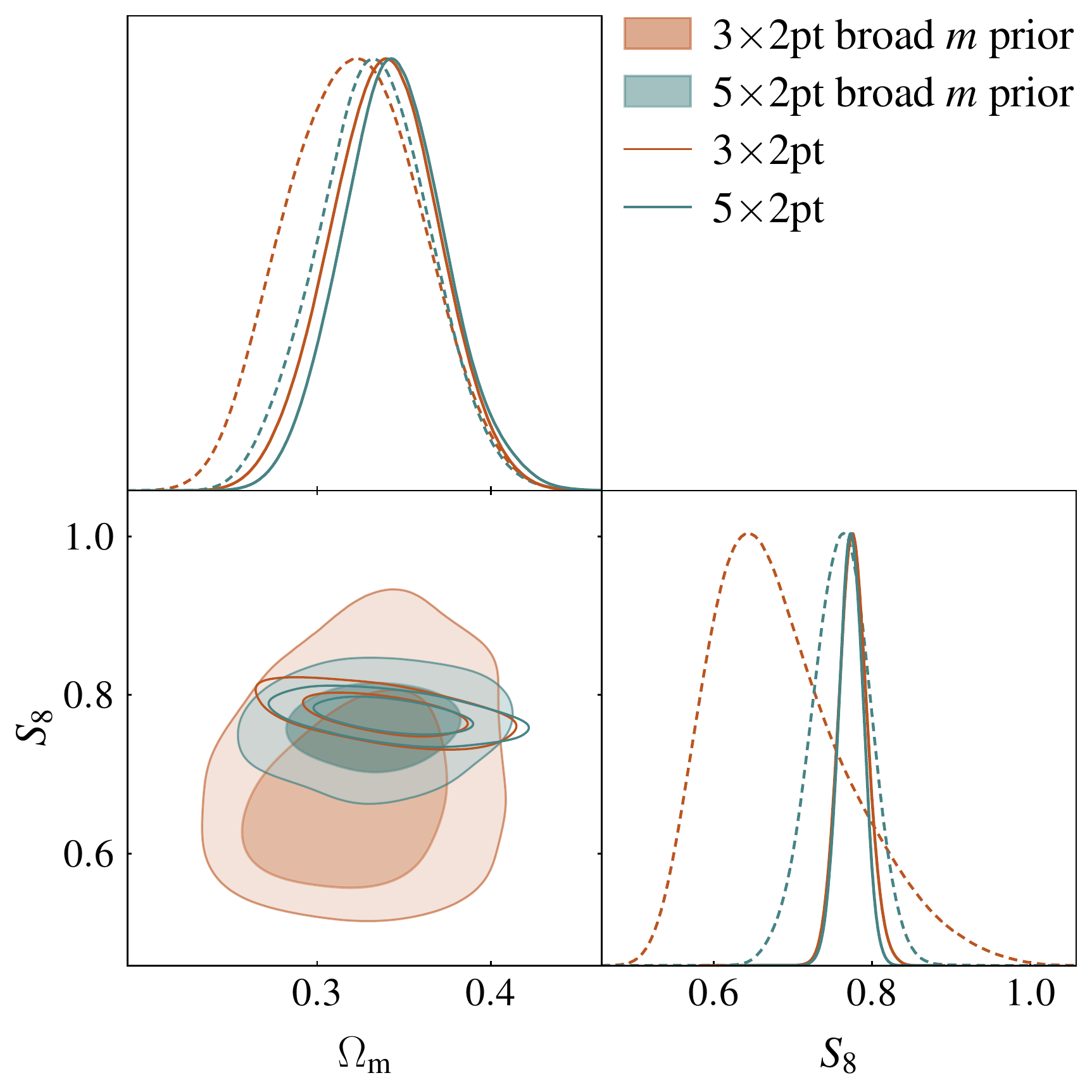}
    \caption{\textit{Left}: Constraints on $S_8$ and the shear calibration parameters, $m_i$, from \threetwo{} and \fivetwo{} using different priors on $m_i$.  With the nominal tight priors on these parameters, \threetwo{} (red dashed) and \fivetwo{} (teal dashed) yield comparable cosmological constraints.  However, when the priors on $m_i$ are substantially weakened, the constraints from \fivetwo{} (teal solid) become significantly tighter than those from \threetwo{} (red solid).  Similarly, the \fivetwo{} analysis obtains tighter constraints on the $m_i$ parameters themselves.  \textit{Right}: Same as left panel, but showing constraints on $S_8$ and $\Omega_{\rm m}$.  Using the broad $m_i$ priors significantly weakens the cosmological constraints from \threetwo{}, but has less of an impact on \fivetwo{}.}
    \label{fig:freem}
\end{figure*}

\subsubsection{Shear calibration}
\label{sec:m_calib}

A potentially significant source of systematic uncertainty impacting cosmological constraints from cosmic shear is biases in shear estimation \citep{Hirata:2003}.  Typically, estimators of lensing shear are calibrated via application to simulated lensed galaxy images.  For the DES Year 3 cosmological analysis, calibration of shear biases is described in \citep{y3-imagesims}.  While this approach can be used to place tight constraints on shear biases, it has the disadvantage of relying on simulated data.  A mismatch between the simulated galaxies used to calibrate the shear estimators and real galaxies could potentially introduce systematic bias.  

As pointed out in \cite{Vallinotto:2012,Baxter:2016,Schaan:2017}, joint analyses of cross-correlations between galaxy surveys and CMB lensing measurements offer the potential of constraining shear calibration biases using only the  data.  To explore this idea, we repeat our analysis of the \threetwo{} and \fivetwo{} data vectors using very wide, flat priors on the shear calibration parameters: $m_i \in (-0.5, 0.5)$.  

The results of this analysis are shown in Fig.~\ref{fig:freem}.  Removing the tight priors on the $m_i$ significantly weakens the cosmological constraints from  \threetwo{}, especially the constraint on $S_8$.  This is because both $m$ and $S_8$ impact the amplitude of the lensing correlation functions, leading to strong degeneracy between the two.  The shear calibration parameters $m_i$ are also very poorly constrained without the tight priors.  However, when the CMB lensing cross-correlations are analyzed jointly with \threetwo{} (i.e. forming \fivetwo{}), the analysis becomes significantly more robust to shear calibration.  Removing the priors on $m_i$ weakens the cosmological constraints, but not nearly as much as for \threetwo{}: Removing the $m$ priors degrades the constraints on $S_8$ by a factor of 4.7 for \threetwo{}, but only by a factor of 2.3 for \fivetwo{} (see right panel of Fig.~\ref{fig:freem}).  The resulting cosmological constraints are consistent with those in the baseline analysis, providing evidence that the DES Y3 \threetwo{} and \fivetwo{} constraints are robust to shear calibration biases.  We also find that the \fivetwo{} data vector achieves constraints on $m$ at roughly the 5--10\% level depending on the redshift bin, roughly a factor of two improvement over the Y1 analysis presented in \citep{5x2Y1}.

\begin{figure*}
	\includegraphics[width=0.9\textwidth]{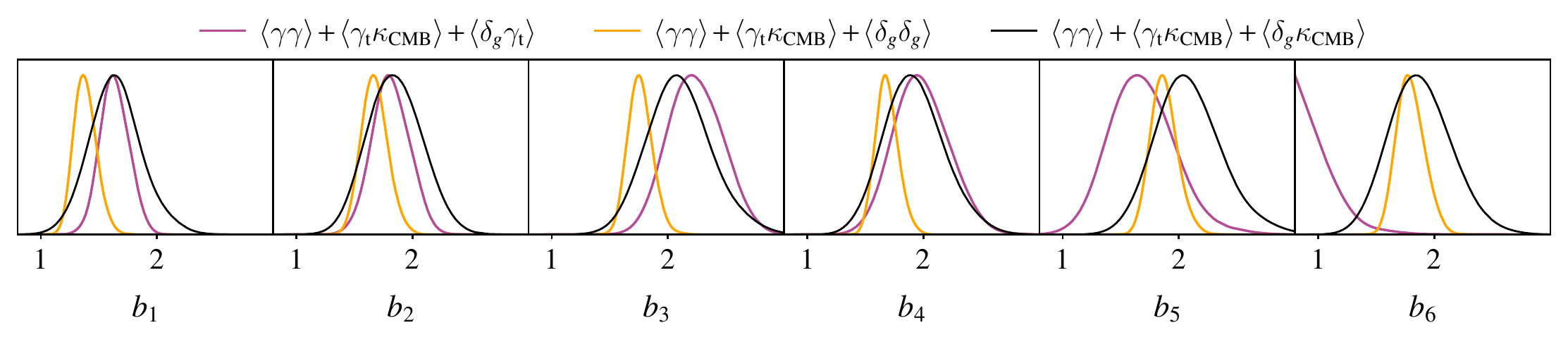}
    \caption{Posteriors on the linear  bias parameters for the \maglim{} galaxies resulting from different combinations of probes.  The parameter $b_i$ represents the linear bias for the $i$th redshift bin.  For the two highest redshift bins (excluded in the baseline cosmology analysis), galaxy clustering (\nn{}) and galaxy-galaxy lensing (\galshear{}) prefer somewhat different values of the bias, with \nk{} more in line with the values preferred by clustering.}
    \label{fig:maglim_X}
\end{figure*} 

\begin{figure*}
	\includegraphics[width=0.75\textwidth]{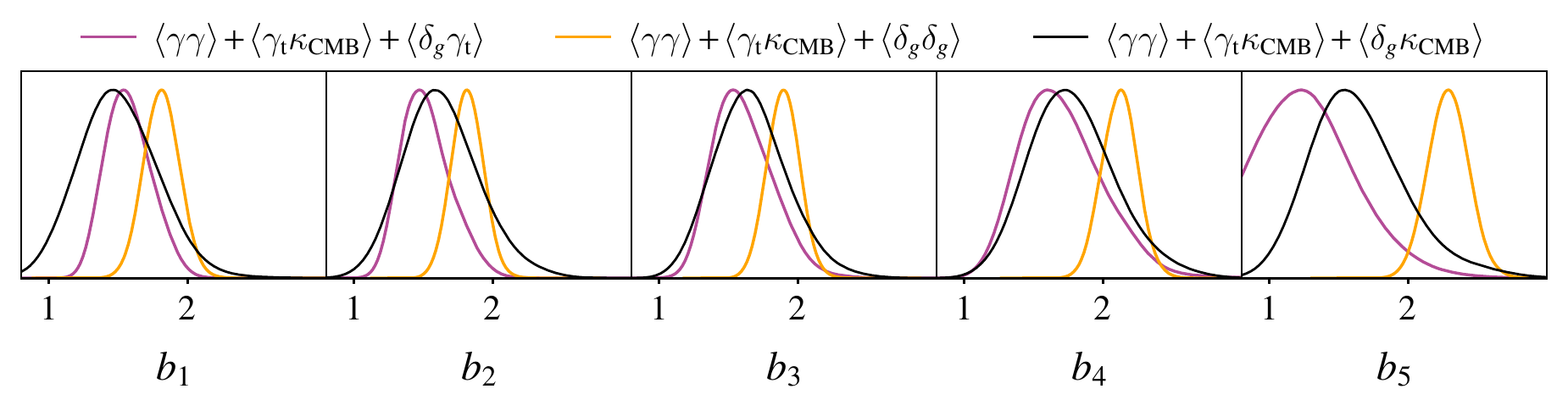}
    \caption{Same as Fig.~\ref{fig:maglim_X}, but for \redmagic{} galaxies.  The bias values preferred by \nk{} are in good agreement with those preferred by \galshear{}, but show a preference for lower bias values than \nn{} across all redshifts.}
    \label{fig:redmagic_X}
\end{figure*}

\subsubsection{Investigating the $X_{\rm lens}$ systematic}
\label{sec:xlens}

As noted previously, analyses of \nn{} and \galshear{} measured with the DES Y3 \maglim{} \cite{y3-2x2ptaltlensresults} and \redmagic{} \cite{y3-2x2ptbiasmodelling} galaxy samples uncovered discrepancies between the values of galaxy bias preferred by these two correlation functions.  The \nn{} measurements with \maglim{} galaxies in the two highest redshift bins (i.e. those shown with the dashed lines in the top panel of Fig.~\ref{fig:nofz}) prefer higher bias values than \galshear{} by roughly 40 to 60\%.  Measurements of \nn{} with the \redmagic{} galaxies, on the other hand, show roughly 10\% higher bias values than the \galshear{} measurements for the first four redshift bins, with this discrepancy increasing to roughly 40\% for the highest redshift bin.  In principle, some difference between the bias values inferred from \nn{} and \galshear{} could result from stochastic biasing \cite[e.g.][]{Baumann:2013}.  However, the amplitude of the difference seen for the \redmagic{} galaxies and the high-redshift \maglim{} galaxies (roughly 10 to 40\% percent) is significantly larger than expected from stochasticity (a few percent) \cite{Desjacques:2018,Pandey:3dbias}.  In \cite{y3-2x2ptbiasmodelling}, a new parameter, $X_{\rm lens}$, was introduced to explore this effect:
\begin{equation}
X_{\rm lens}^i = b^i_{\langle \delta_g \gamma_{\rm t}\rangle} / b^i_{\langle \delta_g \delta_g \rangle},
\end{equation}
where $b^i_{\langle \delta_g \gamma_{\rm t}\rangle}$ ($b^i_{\langle \delta_g \delta_g \rangle}$) is the bias parameter for \galshear{} (\nn{}) in lens galaxy redshift bin $i$.  The finding that some of the clustering measurements prefer a higher value of galaxy bias than the galaxy-galaxy lensing measurements amounts to a preference for $X_{\rm lens}^i < 1$ when we expect $X_{\rm lens}^i = 1$.

The galaxy-CMB lensing cross-correlations also constrain galaxy bias, providing another handle on the anomalous values of the $X_{\rm lens}$ parameter  seen with the \redmagic{} and high-redshift \maglim{} galaxies.   We show constraints on the galaxy bias parameters of the \maglim{} and \redmagic{} galaxies from three combinations of probes in Fig.~\ref{fig:maglim_X} and Fig.~\ref{fig:redmagic_X}, respectively.  Each of the plotted constraints uses the combination of \shearshear{} and \gk{} --- which are effectively independent of the lens galaxies --- to constrain the cosmology.  The remaining probe is then chosen to be \nn{}, \galshear{}, or \nk{}, and this probe is used to constrain the galaxy bias.\footnote{This analysis is similar to that presented in \cite{y3-nkgkmeasurement}, but differs in that we have allowed cosmological parameters to vary, and have included the \shearshear{} and \gk{} measurements (in effect letting the data constrain the cosmological model).}

For the two highest redshift bins of \maglim{} galaxies, we see from Fig.~\ref{fig:maglim_X} that the \nn{} measurements prefer higher values of galaxy bias than the \galshear{} measurements, consistent with the preference for $X_{\rm lens}^i < 1$ described above.  Interestingly, it appears that the \nk{} measurements prefer galaxy bias values more in line with the \nn{} measurements.  This suggest that the preference for  $X_{\rm lens}^i < 1$ is likely driven by \galshear{}.  This is perhaps not surprising, given the large residuals of the model fits to \galshear{} seen in \cite{y3-2x2ptaltlensresults}.  However, note that there is no obvious reason for a possible failure of the baseline model to fit \galshear{}.  As a cross-correlation, the \galshear{} measurements are expected to be quite robust to many observational systematics.  Moreover, any systematic impacting $\delta_{\rm g}$ would likely show up even more strongly in \nn{}, and any systematic impacting $\gamma$ would likely show up more strongly in \shearshear{}.  Another possibility is a failure in modeling some physical effect.  One such effect is lens magnification, which is known to have a significant impact on the \galshear{} correlations at high redshifts \cite{y3-2x2maglimforecast}.

Fig.~\ref{fig:redmagic_X} shows the analogous bias constraints for \redmagic{} galaxies.  In this case, we see that \galshear{} and \nk{} measurements both prefer consistently lower values of galaxy bias than \nn{}, with this difference particularly pronounced in the last redshift bin.  This suggests that a possible cause of the \redmagic{} preference for $X^{\rm lens}_i < 1$ is in the \nn{} measurements.  In the case of \nn{}, it is possible that some observational systematic is modulating the \redmagic{} galaxy overdensity field, resulting in a higher than expected clustering amplitude and thus a preference for higher galaxy bias.  Such a systematic in the $\delta_{\rm g}$ measurements would be expected to have a less noticeable impact on \galshear{}.  At the same time, it should be emphasized that the analysis of \cite{y3-galaxyclustering} extensively tested the \redmagic{} sample for possible contamination by various observational systematics.  While some correlation of known systematics with galaxy density is detected, this correlation is corrected using galaxy re-weighting.  It therefore appears to be difficult to explain the anomalous $X^{\rm lens}$ values with any known observational systematic.

\begin{figure}
	\includegraphics[width=\columnwidth]{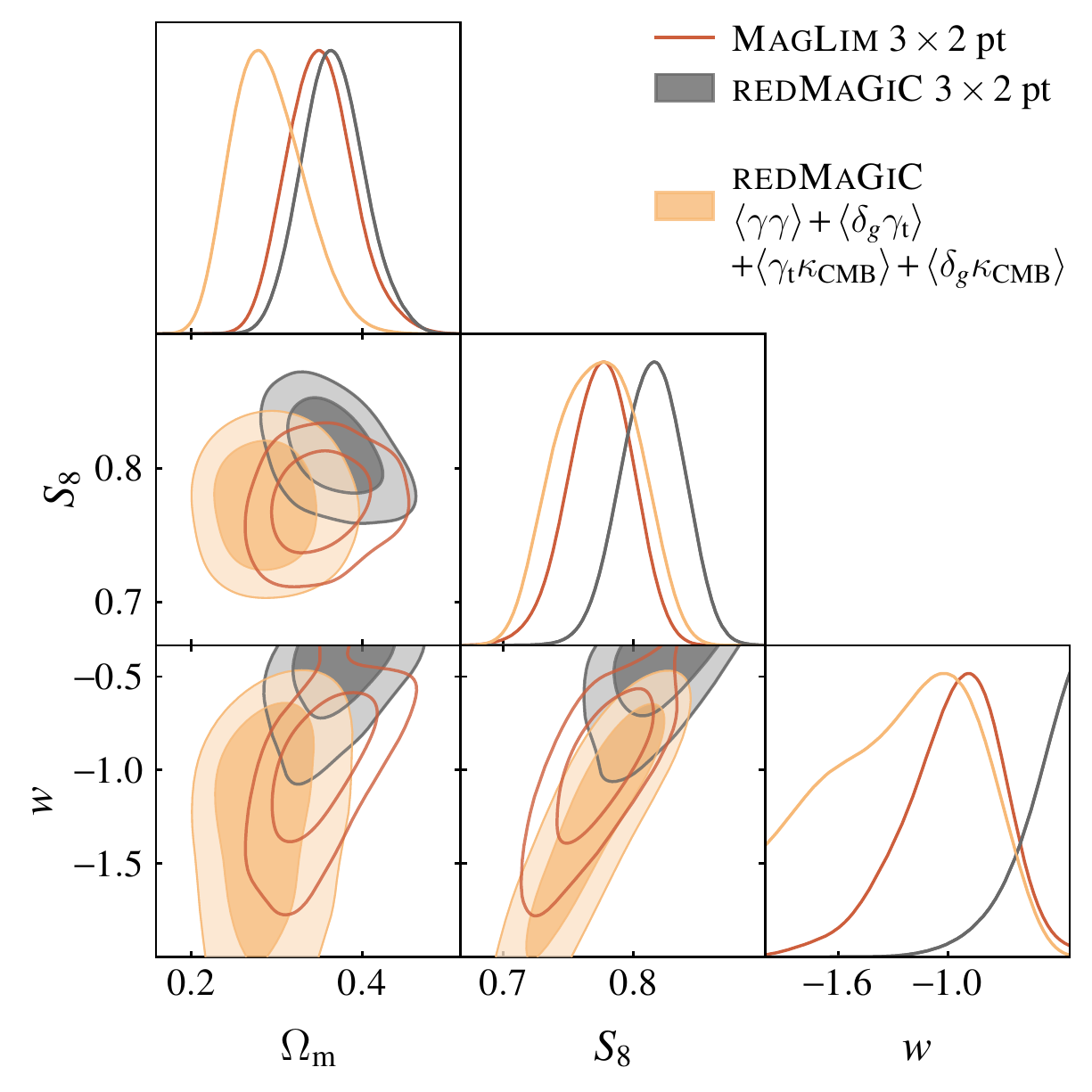}
    \caption{Cosmological constraints on $w$CDM from the \threetwo{} data vector measured with the \maglim{} (red) and \redmagic{} (grey) lens galaxy samples. The constraints from \redmagic{} prefer surprisingly less negative $w$, as discussed in \cite{y3-3x2ptkp}.  However, when the \redmagic{} clustering measurements (\nn{}) are replaced by \nkgk{} to form a combination of four two-point functions (orange), the constraints agree better with those of \maglim{}.
    }
    \label{fig:redmagic_wcdm}
\end{figure}

The interpretation of the \redmagic{} preference for $X^{\rm lens} < 1$ in terms of a systematic impacting \redmagic{} \nn{} measurements is supported by tests with a modified \redmagic{} galaxy sample presented in \cite{y3-2x2ptbiasmodelling}.  The nominal \redmagic{} galaxy sample is selected by requiring that galaxies match a red sequence template, as measured by $\chi^2$.  In \cite{y3-2x2ptbiasmodelling}, an alternative, ``broad $\chi^2$'' sample of galaxies was selected by relaxing the $\chi^2$ threshold for selection.  One would expect that if an observational systematic is modulating the photometry of galaxies, it should have a smaller impact on the ``broad $\chi^2$'' sample than on the nominal sample.  Indeed, it was found that for this alternate sample, the preference for $X_{\rm lens}^i < 1$ seen for the first four redshift bins disappears.  While it might seem surprising that the preference for $X_{\rm lens}^i < 1$ is possibly driven by two different factors for \maglim{} and \redmagic{} galaxies, this interpretation seems consistent with the observed redshift trends.  It may be that observational systematics in \redmagic{} galaxy selection are impacting the bias values inferred from \nn{} at low redshift, while problems in modeling \galshear{} are impacting the bias values inferred for \maglim{} from \galshear{} at high redshift.  The \redmagic{} galaxies may be less affected by this latter systematic, as they do not extend to the high redshifts probed by the last two redshift bins of the \maglim{} sample.  We note, though, that even for \redmagic{}, the CMB lensing cross-correlations prefer higher galaxy bias than \galshear{} in the highest redshift bin; this could be suggesting that the same problem impacting the high-redshift \maglim{} galaxies is impacting the high-redshift \redmagic{} galaxies.  This interpretation would be consistent with mismodeling of \galshear{} at high redshift.

The impact of the apparent systematic in the \redmagic{} sample is also noticeable when the cosmological model is changed from $\Lambda$CDM to $w$CDM.  While the \redmagic{} \threetwo{} constraints on $\Lambda$CDM are quite robust to allowing the $X_{\rm lens}$ parameter to vary, the constraints on $w$CDM shift significantly when this additional freedom is introduced.  This is perhaps not surprising given that the systematic biases with \redmagic{} appear to be redshift-dependent, and might therefore be somewhat degenerate with the effects of $w$.  
\begin{figure*}
	\includegraphics[width=\linewidth]{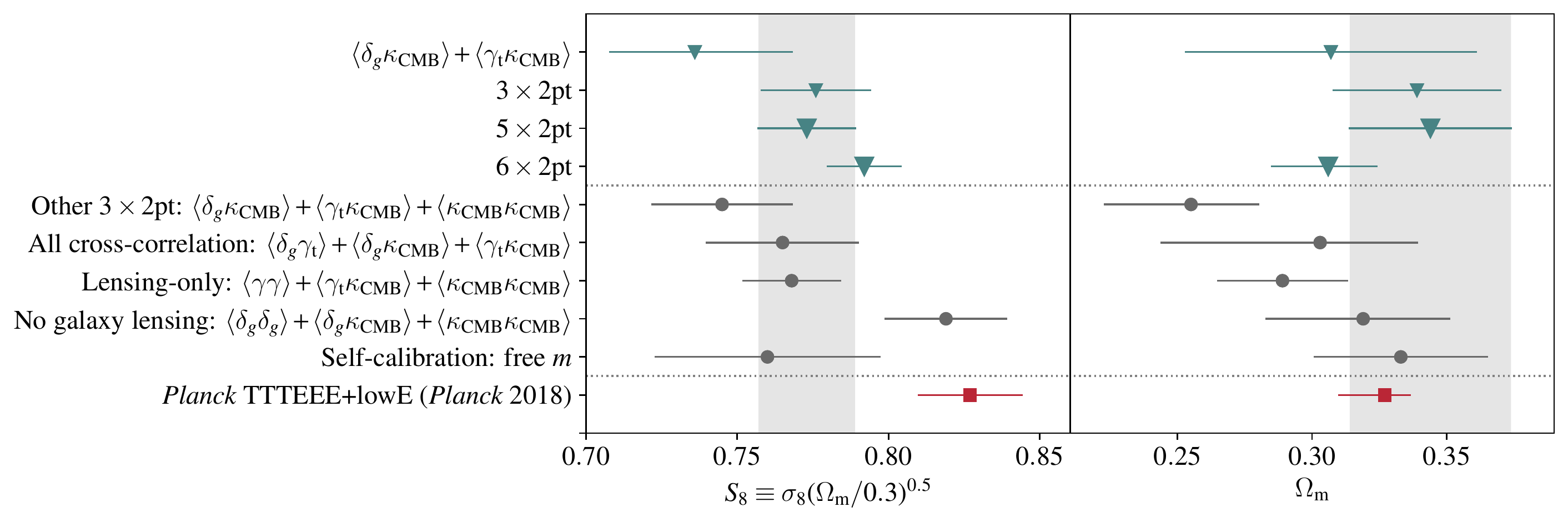}
    \caption{Comparison of the cosmological constraints resulting from different combinations of two-point functions involving DES measurements of galaxy positions and lensing, and SPT+{\it Planck} measurements of CMB lensing.  We also show (bottom row) constraints from {\it Planck}-only measurements of the primary CMB fluctuations.  }
    \label{fig:S8}
\end{figure*} 

Since our analysis above suggests that the problems with \redmagic{} may be isolated to the clustering measurements, in Fig.~\ref{fig:redmagic_wcdm} we present constraints on $w$CDM from the \fivetwo{} combination of probes \textit{without} the clustering measurements.  Interestingly, we see that there is a significant shift in the constraints on $w$ relative to the \threetwo{} analysis.  The constraints without the clustering measurements are in good agreement with the \maglim{} constraints.  This lends additional support to the idea that the \redmagic{} clustering measurements may be systematically biased.  

To summarize the above discussion, our analysis with CMB lensing cross-correlations suggests that there may be two different sources for the $X_{\rm lens}$ systematic seen with \redmagic{} and \maglim{} galaxies.  For \redmagic{} galaxies, our analysis suggests a possible bias in the clustering measurements across all redshift bins.  Such a bias could conceivably be caused by some observational systematic impacting the \redmagic{} selection, which would be consistent with tests performed in \cite{y3-2x2ptbiasmodelling}.  At the same time, high-redshift \maglim{} galaxies (and possibly high-redshift \redmagic{} galaxies as well) show evidence of a potentially different systematic error that favors a problem with the \galshear{} fits.  Such an issue could conceivably be caused by a problem with the \galshear{} modeling, such as an incorrect prescription for magnification effects, which become more pronounced at high redshifts.

\subsection{Consistency with {\it Planck} primary CMB measurements}
\label{sec:tension_planck}

As seen in Fig.~\ref{fig:5x2_kk_6x2}, we find that the cosmological constraints on $\Lambda$CDM from \fivetwo{} and \sixtwo{} are not in significant tension with the constraints from the primary CMB measurements of {\it Planck}.  In particular, we compare our constraints to those from the combination of {\it Planck} $TT$, $TE$, $EE$ and low-$\ell$ $E$-mode polarization measurements ({\it Planck} TT+TE+EE+lowE) \cite{Planck:cosmo}.  Note that we do {\it not} include {\it Planck} measurements of the CMB lensing power spectrum in this combination.  Using the tension metric of \cite{Raveri:2021}, we find that the \threetwo{}, \fivetwo{}, and \sixtwo{} constraints are in agreement with {\it Planck} at the level of $1.5\sigma$, $1.4\sigma$, and $1.4\sigma$, respectively.  The fact that \threetwo{} and \fivetwo{} are roughly equally consistent with {\it Planck} is not surprising, given that the \fivetwo{} constraints are quite close to those of \threetwo{}.  Interestingly, while the  \sixtwo{} constraints are significantly tighter than \fivetwo{}, the level of consistency with {\it Planck} remains roughly the same.   This results from the preference by \kk{} for somewhat higher values of $\sigma_8$, as seen in Fig.~\ref{fig:5x2_kk_6x2}.  Fig.~\ref{fig:S8} directly compares the $S_8$ and $\Omega_{\rm m}$ constraints from these and other two-point function combinations, assuming $\Lambda$CDM.  We note that for consistency with our analysis, we vary the sum of the neutrino masses and impose the priors shown in Table~\ref{table:prior} when generating the {\it Planck} primary CMB constraints shown in this figure.

\section{Summary}
\label{sec:discussion}

We have presented cosmological constraints from an analysis of two-point correlation functions between measurements of galaxy positions and galaxy lensing from DES Y3 data, and CMB lensing measurements from SPT and {\it Planck}.  Our main cosmological constraints are summarized in Table~\ref{tab:param_fiducial}.

The high signal-to-noise of the CMB lensing cross-correlation measurements using DES Y3, SPT-SZ and {\it Planck} data enables powerful robustness tests of our cosmological constraints.  The results of several of these tests are shown in Fig.~\ref{fig:S8}.  We summarize the main findings of these tests below: 
\begin{itemize}
    \item The goodness of fit of $\Lambda$CDM to the \fivetwo{} data vector is acceptable ($p= 0.062$), and the corresponding parameter constraints are consistent with those from \kk{} measurements by {\it Planck}.
    \item Using only cross-correlations between DES and CMB lensing, we obtain constraints on $S_8$ that are comparable in precision and consistent with the baseline \fivetwo{} results.  This result suggests that additive systematics are not significantly impacting the \fivetwo{} cosmological constraints.
    \item Using only gravitational lensing (i.e. no information from galaxy overdensities) yields constraints in agreement with the baseline results.  This result suggests that potential systematics impacting the DES galaxy samples, as well as modeling of galaxy bias, are not significantly biasing the \fivetwo{} cosmological constraints.
    \item The cosmological constraints from two-point functions of \maglim{} galaxy overdensity measurements and CMB lensing are generally consistent with the baseline \fivetwo{} analysis.  This result suggests that shear systematics and modeling of galaxy lensing are not significantly biasing the \fivetwo{} cosmological constraints.  We do, however, observe a low-significance increase in $S_8$ when considering only those two-point functions that do not involve galaxy lensing.  This shift is driven by the intersection of the \nn{} + \nk{} and \kk{} constraints, and is not present when considering \nn{} + \nk{} alone.
    \item Without priors on shear calibration, the cosmological constraints on $S_8$ from \fivetwo{} are in good agreement with the baseline \fivetwo{} results.  The data calibrate the shear bias parameters at the 5--10\% level, and yield constraints consistent with our nominal priors.  These results suggest that shear calibration biases are not significantly impacting the \fivetwo{} cosmological constraints.
    \item The constraints on $\Omega_{\rm m}$ from the different analysis variations are generally consistent.  Although the analysis of \nk{}+\gk{}+\kk{} prefers a somewhat lower value of $\Omega_{\rm m}$, this combination of probes is statistically consistent with \threetwo{}.
\end{itemize}
The cosmological constraints from the \threetwo{}, \fivetwo{}, and \sixtwo{} analyses therefore appear remarkably robust to possible systematic biases.

Assessing the consistency between our constraint on $\Lambda$CDM and those of {\it Planck}, we find that the \fivetwo{} and \sixtwo{} constraints are statistically consistent with {\it Planck} at the $1.4\sigma$ level, as assessed using the full, multi-dimensional posteriors from these measurements. As seen in Fig.~\ref{fig:S8}, however, essentially all combinations of two point functions that we consider prefer lower $S_8$ values than {\it Planck}.  Note, though, that there is significant covariance between some of these measurements.  

We have also investigated possible issues with the analysis of alternate lens galaxy samples, namely the high-redshift \maglim{} galaxies and the \redmagic{} galaxies.  Evidence for biases when analyzing correlation functions measured with these samples was found previously in  \cite{y3-2x2ptaltlensresults}, \cite{y3-2x2ptbiasmodelling}, \cite{y3-2x2ptmagnification}, and \cite{y3-3x2ptkp}.  The CMB lensing cross-correlations considered here provide a powerful way to probe the sources of these biases.  In the context of $\Lambda$CDM, our analysis of CMB lensing cross-correlations suggests a possible problem in the modeling of \galshear{} at high redshift for the \maglim{} galaxies, and possibly the \redmagic{} galaxies as well.  At the same time, the \nk{} measurements with \redmagic{} 
suggest a possible observational systematic that impacts \redmagic{} galaxy clustering across all redshifts.  This interpretation is supported by tests with an alternate \redmagic{} galaxy sample in \cite{y3-2x2ptbiasmodelling}.  In the context of $w$CDM, the \threetwo{} measurements with \redmagic{} have previously shown to yield constraints inconsistent with the \maglim{} analysis, and a preference for surprisingly less negative $w$.  We show that analysis of \shearshear{}+\galshear{}+\nk{}+\gk{} (i.e. two-point functions between DES and CMB lensing, excluding galaxy clustering) measured with \redmagic{} yields cosmological constraints that are in better agreement with \maglim{}, and do not show a strong preference for $w > -1$.  Finally, we note that while the analyses presented here suggest possible interpretations of the $X_{\rm lens}$ bias, more work with current and future DES data is needed to clarify the true source of this systematic uncertainty.

As the data volume and quality from cosmological surveys continue to improve, we expect similar cross-correlation analyses between galaxy surveys and CMB lensing measurements to play an important role in constraining late-time large scale structure.  Excitingly, we expect constraints from such measurements to improve dramatically in the very near future with Year 6 data from DES and new CMB lensing maps from SPT-3G \cite{Sobrin:2022} and AdvACT \cite{Henderson:2016}.  These measurements should help to provide a clearer picture of any possible $S_8$ tension.  Looking farther forward, cross-correlations between surveys such as the Vera Rubin Observatory Legacy Survey of Space and Time \cite{LSST,lsst-science}, the Nancy Grace Roman Space Telescope \cite{Dore2019} , the ESA {\it Euclid} mission \cite{Euclid}, Simons Observatory \cite{SimonsObs}, and CMB-S4 \cite{S4} will  enable significantly more powerful cross-correlation studies that will deliver some of the most precise and accurate cosmological constraints, and that will allow us to continue stress-testing the concordance $\Lambda$CDM model.

\acknowledgements
The South Pole Telescope program is supported by
the National Science Foundation (NSF) through 
the grant OPP-1852617.  Partial support is also 
provided by the Kavli Institute of Cosmological Physics 
at the University of Chicago.
Argonne National Laboratory’s work was supported by
the U.S. Department of Energy, Office of Science, Office of High Energy Physics, under contract DE-AC02-
06CH11357. Work at Fermi National Accelerator Laboratory, a DOE-OS, HEP User Facility managed by the Fermi Research Alliance, LLC, was supported under Contract No. DE-AC02- 07CH11359. The Melbourne authors acknowledge support from the Australian Research Council’s Discovery Projects scheme (DP210102386). The McGill authors acknowledge funding from the Natural Sciences and Engineering Research
Council of Canada, Canadian Institute for Advanced research, and the Fonds de recherche du Qu\'{u}bec Nature et technologies. The CU Boulder group acknowledges support from NSF AST-0956135. The Munich group acknowledges the support by the ORIGINS Cluster (funded by the Deutsche Forschungsgemeinschaft (DFG, German Research Foundation) under Germany’s Excellence Strategy – EXC-2094 – 390783311), the MaxPlanck-Gesellschaft Faculty Fellowship Program, and
the Ludwig-Maximilians-Universit\"{a}t M\"{u}nchen. JV acknowledges support from the Sloan Foundation.

Funding for the DES Projects has been provided by the U.S. Department of Energy, the U.S. National Science Foundation, the Ministry of Science and Education of Spain, 
the Science and Technology Facilities Council of the United Kingdom, the Higher Education Funding Council for England, the National Center for Supercomputing 
Applications at the University of Illinois at Urbana-Champaign, the Kavli Institute of Cosmological Physics at the University of Chicago, 
the Center for Cosmology and Astro-Particle Physics at the Ohio State University,
the Mitchell Institute for Fundamental Physics and Astronomy at Texas A\&M University, Financiadora de Estudos e Projetos, 
Funda{\c c}{\~a}o Carlos Chagas Filho de Amparo {\`a} Pesquisa do Estado do Rio de Janeiro, Conselho Nacional de Desenvolvimento Cient{\'i}fico e Tecnol{\'o}gico and 
the Minist{\'e}rio da Ci{\^e}ncia, Tecnologia e Inova{\c c}{\~a}o, the Deutsche Forschungsgemeinschaft and the Collaborating Institutions in the Dark Energy Survey. 

The Collaborating Institutions are Argonne National Laboratory, the University of California at Santa Cruz, the University of Cambridge, Centro de Investigaciones Energ{\'e}ticas, 
Medioambientales y Tecnol{\'o}gicas-Madrid, the University of Chicago, University College London, the DES-Brazil Consortium, the University of Edinburgh, 
the Eidgen{\"o}ssische Technische Hochschule (ETH) Z{\"u}rich, 
Fermi National Accelerator Laboratory, the University of Illinois at Urbana-Champaign, the Institut de Ci{\`e}ncies de l'Espai (IEEC/CSIC), 
the Institut de F{\'i}sica d'Altes Energies, Lawrence Berkeley National Laboratory, the Ludwig-Maximilians Universit{\"a}t M{\"u}nchen and the associated Excellence Cluster Universe, 
the University of Michigan, NFS's NOIRLab, the University of Nottingham, The Ohio State University, the University of Pennsylvania, the University of Portsmouth, 
SLAC National Accelerator Laboratory, Stanford University, the University of Sussex, Texas A\&M University, and the OzDES Membership Consortium.

Based in part on observations at Cerro Tololo Inter-American Observatory at NSF's NOIRLab (NOIRLab Prop. ID 2012B-0001; PI: J. Frieman), which is managed by the Association of Universities for Research in Astronomy (AURA) under a cooperative agreement with the National Science Foundation.

The DES data management system is supported by the National Science Foundation under Grant Numbers AST-1138766 and AST-1536171.
The DES participants from Spanish institutions are partially supported by MICINN under grants ESP2017-89838, PGC2018-094773, PGC2018-102021, SEV-2016-0588, SEV-2016-0597, and MDM-2015-0509, some of which include ERDF funds from the European Union. IFAE is partially funded by the CERCA program of the Generalitat de Catalunya.
Research leading to these results has received funding from the European Research
Council under the European Union's Seventh Framework Program (FP7/2007-2013) including ERC grant agreements 240672, 291329, and 306478.
We  acknowledge support from the Brazilian Instituto Nacional de Ci\^encia
e Tecnologia (INCT) do e-Universo (CNPq grant 465376/2014-2).

This manuscript has been authored by Fermi Research Alliance, LLC under Contract No. DE-AC02-07CH11359 with the U.S. Department of Energy, Office of Science, Office of High Energy Physics.

We gratefully acknowledge the computing resources provided on Crossover ( and/or Bebop and/or Swing and/or Blues), a high-performance computing cluster operated by the Laboratory Computing Resource Center at Argonne National Laboratory.

\clearpage

\appendix

\section{Parameter priors}
\label{app:priors}

In Table~\ref{table:prior} we list the priors used in our analysis.

\begin{table*}
\centering 
\begin{tabular}{cc}
\hline
Parameter & Prior  \\ \hline
$\Omega_{\rm m} $ & $\mathcal{U}[0.1, 0.9]$  \\
$A_{\rm{s}}\times 10^{9}$ & $\mathcal{U}[0.5, 5.0]$ \\
$n_{\rm{s}}$ & $\mathcal{U}[0.87, 1.07]$ \\
$\Omega_{\rm{b}}$ & $\mathcal{U}[0.03, 0.07]$  \\
$h$ & $\mathcal{U}[0.55, 0.91]$\\
$\Omega_{\nu}h^2 \times 10^{4} $& $\mathcal{U}[6.0, 64.4]$  \\  
\hline
$a_1$ & $\mathcal{U}[-5.0, 5.0]$ \\
$a_2$ & $\mathcal{U}[-5.0, 5.0]$\\
$\eta_1$ & $\mathcal{U}[-5.0, 5.0]$ \\
$\eta_2$ & $\mathcal{U}[-5.0, 5.0]$ \\
$b_{\rm{ta}}$ & $\mathcal{U}[0.0, 2.0]$ \\ 
\hline
\textsc{MagLim} &  \\
$b^{1\cdots 6}$  & $\mathcal{U}[0.8, 3.0]$ \\
$b_{1}^{1\cdots 6}$  & $\mathcal{U}[0.67, 3.0]$\\
$b_{2}^{1\cdots 6}$ & $\mathcal{U}[-4.2, 4.2]$ \\
$C_{\rm l}^{1\cdots 6}$ & $\delta(0.42)$, $\delta(0.3)$, $\delta(1.76)$, $\delta(1.94)$, \textcolor{gray}{$\delta(1.56)$, $\delta(2.96)$}  \\
$\Delta_z^{1...6} \times 10^{2}$ & $\mathcal{N}[-0.9, 0.7]$, $\mathcal{N}[-3.5, 1.1]$, $\mathcal{N}[-0.5, 0.6]$,  $\mathcal{N}[-0.7, 0.6]$, \textcolor{gray}{$\mathcal{N}[0.2, 0.7]$ ,  $\mathcal{N}[0.2, 0.8]$}  \\
$\sigma_{z}^{1...6}$ & $\mathcal{N}[0.98, 0.062]$, $\mathcal{N}[1.31, 0.093]$, $\mathcal{N}[0.87, 0.054]$, $\mathcal{N}[0.92,  0.05]$, \textcolor{gray}{$\mathcal{N}[1.08, 0.067]$, $\mathcal{N}[0.845, 0.073]$}  \\
 \hline
\textcolor{gray}{\redmagic{}} &  \\
\textcolor{gray}{$b^{1\cdots 5}$}  & \textcolor{gray}{$\mathcal{U}[0.8, 3.0]$} \\
\textcolor{gray}{$b_{1}^{1\cdots 5}$}  & \textcolor{gray}{$\mathcal{U}[0.67, 2.52]$} \\
\textcolor{gray}{$b_{2}^{1\cdots 5}$} & \textcolor{gray}{$\mathcal{U}[-3.5, 3.5]$} \\
\textcolor{gray}{$C_{\rm l}^{1\cdots 5}$} & 
\textcolor{gray}{$\delta(0.62)$, $\delta(-3.04)$, $\delta(-1.32)$, $\delta(2.5)$, $\delta(1.94)$} \\
\textcolor{gray}{$\Delta_z^{1...5} \times 10^{2}$} & \textcolor{gray}{$\mathcal{N}[0.6, 0.4]$, $\mathcal{N}[0.1, 0.3]$, $\mathcal{N}[0.4, 0.3]$,}  \textcolor{gray}{$\mathcal{N}[-0.2, 0.5]$, $\mathcal{N}[-0.7, 1.0]$}  \\
\textcolor{gray}{$\sigma_{z}^{1...4}$} & \textcolor{gray}{$\delta(1)$, $\delta(1)$, $\delta(1)$, $\delta(1)$, $\mathcal{N}[1.23, 0.054]$}  \\
\hline
\textsc{MetaCalibration} &  \\
$m^{1...4} \times 10^{3}$ &  $\mathcal{N}[-6.0, 9.1]$, $\mathcal{N}[-20.0, 7.8]$,  $\mathcal{N}[-24.0, 7.6]$, $\mathcal{N}[-37.0, 7.6]$ \\
$\Delta_z^{1...4} \times 10^{-2}$ & $\mathcal{N}[0.0, 1.8]$, $\mathcal{N}[0.0, 1.5]$,  $\mathcal{N}[0.0, 1.1]$, $\mathcal{N}[0.0, 1.7]$   \\
\hline
\end{tabular}
\caption{Prior values for cosmological and nuisance parameters included in our model. For the priors, $\mathcal{U}[a,b]$ indicates a uniform prior between $a$ and $b$, while $\mathcal{N}[a,b]$ indicates a Gaussian prior with mean $a$ and standard deviation $b$. $\delta(a)$ is a Dirac Delta function at value $a$, which effectively means that the parameter is fixed at $a$. Note that the fiducial lens sample is the first 4 bins of the \maglim{} sample. The two high-redshift \maglim{} bins and the \redmagic{} sample are shown in grey to indicate they are not part of the fiducial analysis.}
\label{table:prior}
\end{table*}

\section{Adding small-scale information with nonlinear galaxy bias}
\label{app:nonlinear_bias}

Our baseline analysis adopts a linear galaxy bias model to describe the relationship between the galaxy overdensity and the underlying matter field.  At small scales, this description of galaxy biasing is known to break down.  The breakdown in linear galaxy bias drives our choice of angular scales used to analyzing the \nk{} correlation, as described in \citetalias{y3-nkgkmethods}.  By adopting a higher-order bias model, it is possible to include smaller angular scales in the cosmological analysis and potentially improve parameter constraints.  At the same time, a more complex bias model necessitates more free parameters, which degrades the parameter constraints to some extent.  We now consider the parameter constraints from \fivetwo{} using the nonlinear galaxy bias model described in \cite{y3-2x2ptbiasmodelling}.

\begin{figure}
	\includegraphics[width=\columnwidth]{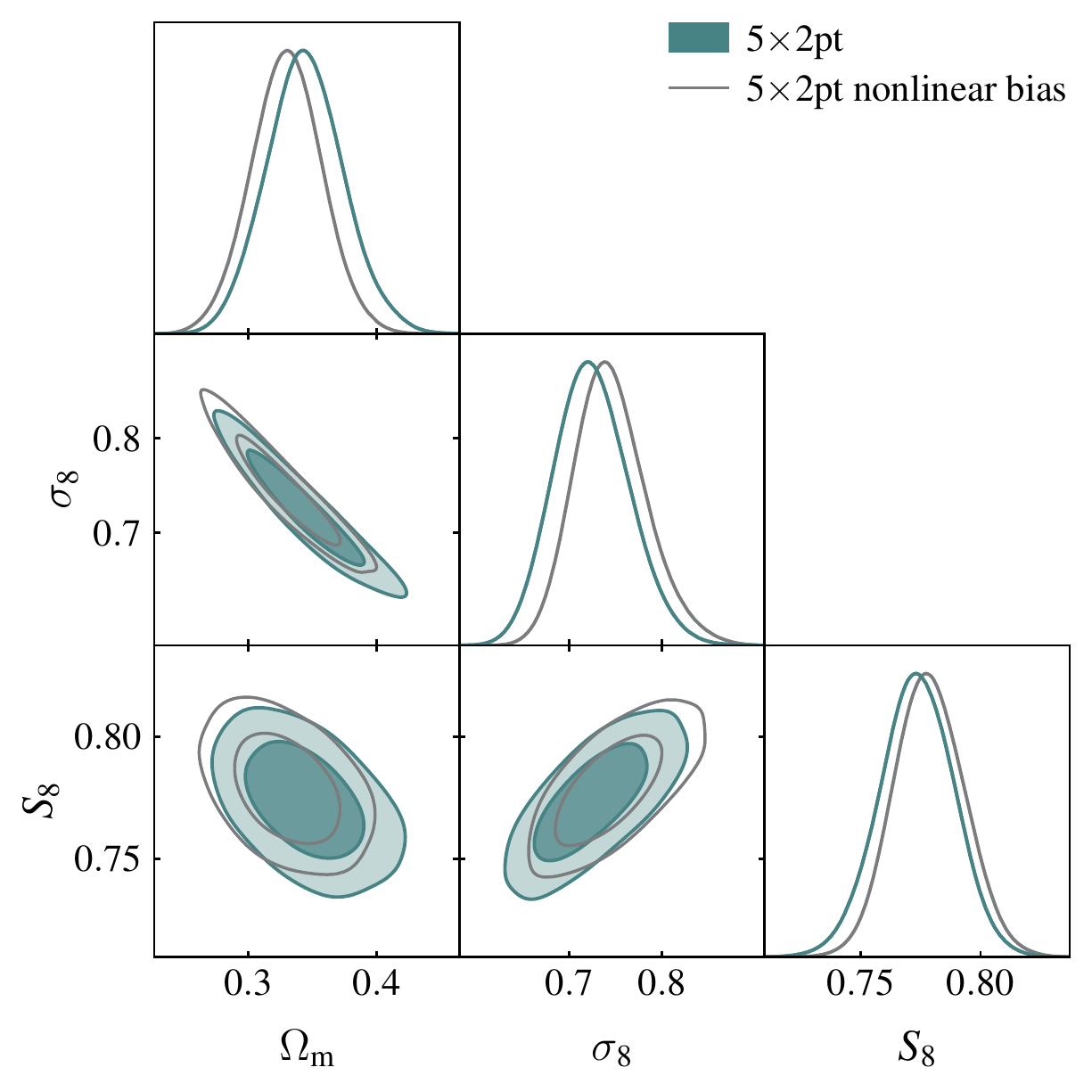}
    \caption{Parameter constraints obtained when using a nonlinear galaxy bias model to analyze the \fivetwo{} data vector (grey) compared to our baseline \fivetwo{} analysis (teal), which adopts a linear bias model.  The nonlinear bias analysis can be used to fit smaller scales of measured correlation functions resulting in improved constraints.}
    \label{fig:nonlinear_bias}
\end{figure}

The constraints from this analysis are presented in Fig.~\ref{fig:nonlinear_bias}.  We find that adopting a nonlinear description of galaxy bias improves the precision of the constraint on $S_8$ by roughly 10\%, and the precision of the constraints on both $\Omega_m$ and $S_8$ by roughly 10\%.

\begin{figure}
	\includegraphics[width=\columnwidth]{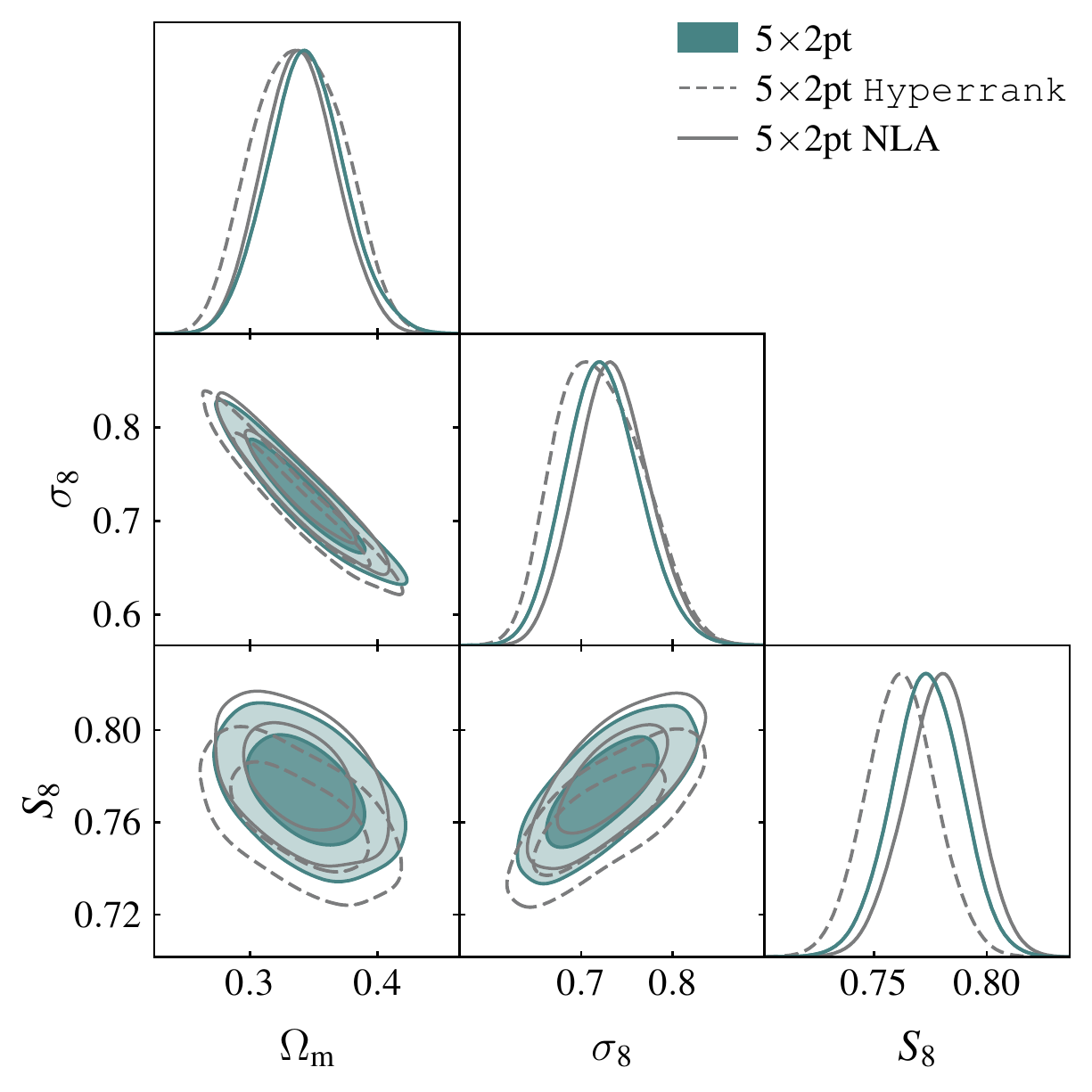}
    \caption{Parameter constraints obtained when using alternative prescriptions for modeling photometric redshift biases and intrinsic alignments.  The teal curves show our baseline results, while the grey dashed curves show results assuming the \textsc{hyperrank} method for calibrating the source galaxy redshift distributions, and the grey solid curves show results assuming the NLA intrinsic alignment model (rather than the baseline TATT model).  In both cases, there are minimal shifts relative to our baseline results.}
    \label{fig:photoz_IA}
\end{figure}

\section{Alternative redshift calibration and IA model}
\label{app:photoz_IA}

Our baseline analysis assumes that uncertainties in the source galaxy redshift distributions are characterized by shift and stretch parameters, as described in \citep{y3-sompz, y3-sourcewz}.  An alternative approach to characterizing the uncertainties in the redshift distributions is \textsc{hyperrank}, described in \cite{y3-hyperrank}.  Rather than attempt to parameterize biases in the redshift distributions, \textsc{hyperrank} provides a way to sample over realizations of the full posteriors on these distributions.  Repeating our analysis of the \fivetwo{} data using this alternative redshift uncertainty prescription yields the constraints shown in Fig.~\ref{fig:photoz_IA}.  Although there is a small shift in $S_8$, it is well within our uncertainties.

The intrinsic alignment (IA) model that we adopt in our baseline analysis is TATT \citep[TATT,][]{blazek2019}.  In Fig.~\ref{fig:photoz_IA}, we show the results of instead adopting the nonlinear alignment model \citep[NLA,][]{bridle07}.  The NLA model is more restrictive than TATT in the sense that the latter becomes equivalent to the former in the limit that $a_2=\eta_2=b_{\rm ta}=0$.  We find that switching to NLA results in minimal changes to the parameter constraints from \fivetwo{}.

\newpage

\bibliography{thebibliography,y3kp}

\label{lastpage}

\end{document}